\documentclass[fleqn,useAMS,usenatbib]{mnras}

\usepackage{graphicx}	% Including figure files
\usepackage{amsmath}	% Advanced maths commands
\usepackage{amssymb}	% Extra maths symbols
\usepackage{multicol}        % Multi-column entries in tables
\usepackage{bm}		% Bold maths symbols, including upright Greek
\usepackage{pdflscape}	% Landscape pages
\usepackage{natbib}
\usepackage{amsmath,breqn} 
\usepackage[T1]{fontenc}
\usepackage{aecompl}

\title{\mbox Study of X-ray variability and coronae of Seyfert galaxies using {\it NuSTAR}}

\author[Priyanka Rani et al.] {
Priyanka Rani$^{1}$\thanks{E-mail: priyanka@iiap.res.in}, 
C. S. Stalin$^{1}$ and
K. D. Goswami$^{2}$
\\
$^{1}$Indian Institute of Astrophysics, Koramangala, Bangalore 560034, India. \\
$^{2}$Department of Physics, Dibrugarh University, Dibrugarh 786004, Assam, India. \\
}

\begin{document}
\label{firstpage}
\pagerange{\pageref{firstpage}--\pageref{lastpage}}
\maketitle

%%%---------------------Abstract-------------------------%%%%
\begin{abstract}
The primary X-ray emission in active galactic nuclei (AGN) originates in a 
compact region  called the corona located very close to the super-massive 
black hole and the accretion disk. The knowledge of the cut-off 
energy ($E_{cut}$) of the primary X-ray continuum in an AGN is very important 
as it carries information on the physical characteristics of the hot 
X-ray emitting corona. 
We present here the results
of our investigation on the spectral properties of a sample of 10 nearby 
AGN (0.005 $<$ $z$ $<$ 0.037) observed with the {\it Nuclear Spectroscopic
Telescope Array (NuSTAR)} . From fitting  of the {\it NuSTAR} data
of 10 sources, we derived clear $E_{cut}$ values 
for the  first time in 8 sources and a lower limit in one source, thereby, doubling the number of AGN with 
$E_{cut}$ measurements from {\it NuSTAR} data. 
Broad Fe K$\alpha$ line was noticed in 7 sources, while, excess emission in 
the energy range beyond $\sim$15 keV arising from Compton reflection was seen in 
all the  sources. We also investigated the correlation 
of $E_{cut}$ with various 
physical characteristics of the AGN such as black hole mass ($M_{BH}$), 
Eddington ratio ($\lambda_{Edd.}$) and X-ray photon index ($\Gamma$). We found no correlation 
between $E_{cut}$ and $M_{BH}$ and between $E_{cut}$ and $\lambda_{Edd.}$, however, $E_{cut}$
correlates with $\Gamma$ in a complex manner. Also, timing 
analysis of the 10 sources indicates that they all are variable with indications 
of more variations in the 
soft band relative to the hard band in some individual sources, however, considering 
all the sources together, the
variations are indistinguishable between hard and soft bands.
 
\end{abstract}


%%%%%-----------------Introductory section-----------------%%%%%
\section{Introduction}
Active Galactic Nuclei (AGN) are the most powerful  
and persistent 
extragalactic sources that emit radiation over a wide range of 
the accessible wavelengths from low energy radio to high energy gamma rays. 
Their luminosities range from 10$^{40}$ erg s$^{-1}$ to larger than
10$^{47}$ erg s$^{-1}$ \citep{1999PNAS...96.4749F}.
They are believed to be powered by accretion of matter on to super-massive black 
holes (SMBHs) located at the centres of galaxies \citep{1984ARA&A..22..471R}. The 
process of accretion leads to the
formation of  an optically thick, geometrically thin accretion disk 
\citep{1973A&A....24..337S}
around the 
SMBH. Such an accretion disk with temperatures in the range of
10$^4$ $-$ 10$^5$ K for a black hole of mass 10$^6$ $-$ 10$^9$ M$_{\odot}$
emits primarily in 
the optical/UV region of the electromagnetic spectrum and the observed
big blue bump (BBB) in the broad band spectral energy distribution of an AGN is 
a signature of the accretion disk. The observed thermal emission from the 
standard accretion disk is a combination of several black bodies of different
temperatures and it depends on the size of the emitting region, the accretion
rate and the mass of the central SMBH. On the other hand, the observed primary
X-ray continuum emission from AGN is thought to be due to the inverse Compton 
scattering of UV and optical photons from the accretion disk by hot electrons
in a compact region called the 
corona \citep{1991ApJ...380L..51H,1994ApJ...432L..95H,1997ApJ...476..620H}. 
This inverse Compton scattering of UV/optical photons from the 
accretion disk by the coronal electrons produces a X-ray continuum with 
a power-law spectral shape and a high energy cut-off \citep{1979rpa..book.....R}. 
This high energy cut-off (E$_{cut}$) in the observed spectrum happens when the electrons
in the corona are no longer able to add energy to the photon in the photon-electron
interaction \citep{2018MNRAS.481.4419B}.
The shape of the power
law continuum depends on various parameters such as the seed photon field,
the coronal temperature ($K_BT_e$), the optical depth and the observers viewing
angle, while the value of $E_{cut}$ is determined by $K_BT_e$ \citep{1993ARA&A..31..717M}. 

\begin{table*}
\caption{\label{log}Details of the sources analysed in this work. The columns are:  
(1) running number,(2) name of the source,  (3) right ascension in hour:minute:seconds, (4) declination in degree:arcminute:arcseconds, (5)
redshift, (6) V-band magnitude, (7) type of the source, (8) Observational IDs, (9) date of observation and
(10) the exposure time in seconds. The values of  $\alpha_{2000}$, $\delta_{2000}$, $z$, V-band magnitude and type of the source were taken from  \citet{2010A&A...518A..10V} catalog.}
\centering
\begin{tabular}{@{}lllllllllr@{}}
\hline
No.  & Name	  &   $\alpha_{2000}$   &	$\delta_{2000}$  & $z$    &  V (mag)  &   Type    & OBSID  & Date  & Exposure \\
(1)  &  (2)       & (3)                 & (4)                    & (5)    & (6)   & (7)      & (8)    & (9)   & (10)   \\
\hline 
1. &  Mrk 348	  & 00:48:47.2    & +31:57:25.0      & 0.014  & 14.59    & Sy1h    & 60160026002  & 2015-10-28  & 21520 \\
2. & Mrk 1040	  & 02:28:14.4    & +31:18:41.0      & 0.016  & 14.74    & Sy1     & 60101002002  & 2015-08-12  & 62960 \\	
   &              &               &                  &        &          &         & 60101002004  & 2015-08-15  & 64252 \\
3. & ESO 362$-$G18& 05:19:35.8    & -32:39:27.0      & 0.013  & 13.37    & Sy1.5   & 60201046002  & 2016-09-24  & 101906 \\
4. & NGC 2992	  & 09:45:42.0    & -14:19:35.0      & 0.008  & 13.78    & Sy1.9   & 60160371002  & 2015-12-02  & 20798  \\	
5. & NGC 3783	  & 11:39:01.8    & -37:44:19.0      & 0.009  & 13.43    & Sy1.5   & 60101110002  & 2016-08-22  & 41271  \\	
   &              &               &                  &        &          &         & 60101110004  & 2016-08-24  & 42434 \\
6. & 4U 1344$-$60 & 13:47:36.0    & -60:37:03.0      & 0.013  & 19.00    & Sy1     & 60201041002  & 2016-09-17  & 99464    \\	
7. & ESO141-G55	  & 19:21:14.3    & -58:40:13.0      & 0.037  & 13.64    & Sy1.2   & 60201042002  & 2016-07-15  & 93011 \\	
8. & Mrk 509	  & 20:44:09.7    & -10:43:24.0      & 0.035  & 13.12    & Sy1.5   & 60101043002  & 2015-04-29  & 165893 \\
   &              &               &                  &        &          &         & 60101043004  & 2015-06-02  & 36475 \\
9. & NGC 7172	  & 22:02:01.9    & -31:52:08.0      & 0.009  &   13.61  & Sy2     & 60061308002  & 2014-10-07  & 32001 \\
10.& NGC 7314	  & 22:35:46.1    & -26:03:02.0      & 0.005  &   13.11  & Sy1h    & 60201031002  & 2016-05-13  & 100424 \\
\hline 
\end{tabular}
\end{table*}

\begin{figure}
\hspace*{-0.5cm}\includegraphics[scale=0.6]{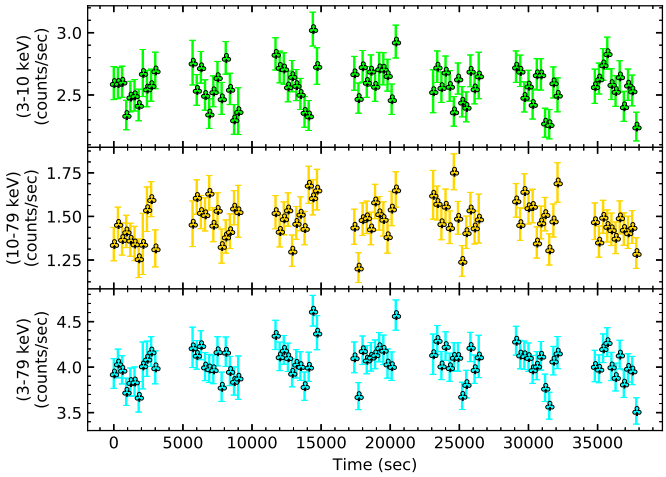}
\caption{\label{Fig-1}Light curves for Mrk 348 in 3$-$10 keV (top panel), 10$-$79 keV (middle panel) and 3$-$79 keV (bottom panel)}
\end{figure}

X-ray reverberation studies indicate 
the corona to be located close to the accretion disk 
\citep{2009Natur.459..540F,2013MNRAS.434.1129K} typically within 3 $-$ 10 $R_G$, 
where, $R_G$ is the gravitational
radius defined as $R_G = GM_{BH}/c^2$, and M$_{BH}$ is the mass of the SMBH. Rapid X-ray
flux variability studies \citep{2005MNRAS.359.1469M}, the observed small 
time scales of X-ray eclipses \citep{2005ApJ...623L..93R,2011MNRAS.410.1027R} 
and 
micro-lensing studies \citep{2009ApJ...693..174C} point to the small size of the X-ray corona
of 5$-$10 $R_G$. Exchange of energy happens between particles and photons in the 
compact corona, the compactness of which is parameterised by the 
dimensionless compactness parameter \citep{1983MNRAS.205..593G}
\begin{equation}
l = 4 \pi \frac{m_p}{m_e} \frac{R_G}{R} \frac{L}{L_E}
\end{equation}

where $m_p$ and m$_e$ are the masses of the proton and electron 
respectively, $R_G$ is the gravitational radius, $R$ is the size of the
source, $L$ is the luminosity of the source, and $L_E$ is the 
Eddington luminosity defined as $L_{Edd} = \frac{4 \pi G M_{BH} m_p c}{\sigma_T} \sim
1.3 \times 10^{38} (\frac{M_{BH}}{M_{\odot}})$ erg s$^{-1}$, where $\sigma_T$ is the
Thomson scattering cross section.  Also, the coronal electron temperature 
can be characterised by the dimensionless parameter as 
\begin{equation}
\theta = K_B T_e/m_e c^2. 
\end{equation}

Empirically, the electron
temperature and E$_{cut}$ are related as 
$E_{cut} = 2 - 3 K_B T_e$ \citep{2001ApJ...556..716P}. 
In spite of several studies, the geometry of the corona is still not known. It 
could be in the form of a  slab \citep{1997ESASP.382..401P} or  
a sphere \citep{1997ApJ...487..759D}. Also, it is not known if
the medium of the coronal plasma is continuous or  patchy 
\citep{1995ApJ...449L..13S,2013A&A...549A..73P}. In addition to the BBB and
the primary power law X-ray continuum with a high energy cut off, a large fraction of  
AGN also show soft excess between 0.1 $-$ 2 keV,  broad ($\sim$ 4 $-$ 7 keV) Fe K$\alpha$ line  and
a Compton reflection bump peaking between 20 $-$ 30 keV. These features are well 
explained by reflection models \citep{2002MNRAS.331L..35F} where the coronal photons 
irradiate the accretion disk and Compton scatter off the disk.

\begin{table*}
\caption{\label{log-variability}Results of variability analysis: Columns are (1) number, (2) name of the sources, (3) observational ID, (4), 
(5) and (6) are the 
$F_{var}$ and error in $F_{var}$ values in the soft (3$-$10 keV), hard (10$-$79 keV) and total (3$-$79 keV) bands respectively.}
\centering
\begin{tabular}{@{}rlllll@{}}
\hline 
    & 		&		        & \multicolumn{3}{c}{$F_{var} \pm F_{var}^{err}$}     \\
No. & Name           &       OBSID           & 3$-$10 keV           & 10$-$ 79 keV          & 3$-$79 keV   \\ 
(1) &  (2)           &   (3)            &  (4)                      & (5)                   & (6)          \\ \hline

1.  &  Mrk 348		&	60160026002	&  0.029$\pm$0.006	&  0.029$\pm$0.008	&  0.026$\pm$0.005 \\
2.  & Mrk 1040	&	60101002002	&  0.089$\pm$0.005	&  0.069$\pm$0.007	&  0.081$\pm$0.004 \\
    &		&	60101002004	&  0.099$\pm$0.005	&  0.070$\pm$0.008	&  0.090$\pm$0.004 \\
3.  & ESO 362$-$G18	&	60201046002	&  0.222$\pm$0.006	&  0.150$\pm$0.008	&  0.195$\pm$0.005 \\
4.  & NGC 2992	&	60160371002	&  0.069$\pm$0.005	&  0.064$\pm$0.008	&  0.067$\pm$0.004 \\
5.  & NGC 3783	&	60101110002	&  0.097$\pm$0.006	&  0.041$\pm$0.009	&  0.093$\pm$0.005 \\
    &		&	60101110004	&  0.057$\pm$0.005	&  0.017$\pm$0.006	&  0.044$\pm$0.004 \\
6.  & 4U 1344$-$60	&	60201041002	&  0.105$\pm$0.003	&  0.094$\pm$0.004	&  0.101$\pm$0.002 \\
7.  & ESO141G055	&	60201042002	&  0.099$\pm$0.004	&  0.075$\pm$0.006	&  0.094$\pm$0.003 \\
8.  & Mrk 509		&	60101043002	&  0.033$\pm$0.002	&  0.044$\pm$0.003	&  0.032$\pm$0.002 \\
    &		&	60101043004	&  0.044$\pm$0.005	&  0.085$\pm$0.008	&  0.049$\pm$0.004 \\
9.  & NGC 7172	&	60061308002	&  0.071$\pm$0.005	&  0.071$\pm$0.006	&  0.077$\pm$0.004 \\
10. &NGC 7314	&	60201031002	&  0.271$\pm$0.003	&  0.195$\pm$0.005	&  0.250$\pm$0.003 \\ 
\hline 

\end{tabular}
\end{table*}

Determination of $E_{cut}$ values from the X-ray spectra of AGN can provide important 
constraints on the temperature of the corona $K_BT_e$. However, $E_{cut}$ measurements are
difficult to obtain due to the requirement of X-ray data with high S/N beyond 10 keV.
In spite of that, $E_{cut}$ measurements for few nearby AGN are available  via analysis of
data from the {\it Compton Gamma Ray Observatory} (CGRO; \citealt{2000ApJ...542..703Z,1996MNRAS.283..193Z,1997ApJ...482..173J}), BeppoSAX \citep{2000ApJ...536..718N,2007A&A...461.1209D},
{\it INTEGRAL} \citep{2014ApJ...782L..25M,2010MNRAS.408.1851L,2016MNRAS.458.2454L,
2011A&A...532A.102R}, Swift/BAT \citep{2013ApJ...763..111V,2017ApJS..233...17R} and Suzaku 
\citep{2011ApJ...738...70T}.
The launch of the {\it Nuclear Spectroscopic Telescope Array} (NuSTAR; \citealt{2013ApJ...770..103H})  with its  unique capability to focus hard X-rays 
and thereby providing good signal to noise ratio data in the 3$-$79 keV band has allowed
us to get improved values of $E_{cut}$ measurements for a growing number of AGN. As 
of  today $E_{cut}$ has been measured in less than two dozen AGN using {\it NuSTAR}  
\citep{2014ApJ...781...83B, 2015ApJ...800...62B,2014ApJ...794...62B,
2015MNRAS.447.3029M,2015MNRAS.452.3266U, 2015ApJ...814...24L,2016arXiv161205871T,
2016A&A...590A..77L,2017arXiv170403673L,2017arXiv170309815K,
2018MNRAS.473.3104T,2018A&A...609A..42P, 2018MNRAS.481.4419B,2018JApA...39...15R,
2018ApJ...856..120R}. With the availability of more spectral measurements from 
{\it NuSTAR}, correlations between $E_{cut}$ and various physical properties of 
the sources have been explored. From an analysis of 12 sources observed with {\it NuSTAR}, 
\cite{2018ApJ...856..120R}, found no correlation of $E_{cut}$ with $M_{BH}$ and
$\Gamma$. Also,  \cite{2018A&A...614A..37T} using 19 sources from {\it NuSTAR}, found
no correlation between $E_{cut}$ and $M_{BH}$. The number of $E_{cut}$
measurements from {\it NuSTAR} is small to unambiguously know for the existence of correlation of $E_{cut}$
with various physical properties of AGN. It is therefore very important to determine
$E_{cut}$ for more number of AGN. Recently, using Swift/BAT data for 
a sample of about 200  AGN, \cite{2018MNRAS.480.1819R} found a statistically significant
negative correlation between $E_{cut}$ and Eddington ratio. However, observations
from Swift/BAT are likely to be background dominated due to its survey mode of operation
necessitating more $E_{cut}$ measurements from {\it NuSTAR} to confirm the correlation noticed by
\cite{2018MNRAS.480.1819R}.
 
To increase the number of AGN that have  $E_{cut}$ measurements, we are 
carrying out a systematic analysis of the spectra of few bright AGN observed with 
{\it NuSTAR}.
In this paper we present the first time measurements of $E_{cut}$ values for 10 AGN. In
section 2 we give our sample and outline the reduction procedures. In Section 3, we 
describe the model fits to the extracted spectra as well as explore the existence of
correlation between $E_{cut}$ and various physical properties of AGN using an updated list
of $E_{cut}$ measurements from {\it NuSTAR} and summarize our findings in the final section. 
Through out the paper, we adopt a cosmology with $H_0$ = 71 km s$^{-1}$ Mpc$^{-1}$, $\Omega_A$ = 0.73 and $\Omega_m$ = 0.27.

\begin{figure*}
\vbox{
      \hbox{
            \hspace*{-0.5cm}\includegraphics[scale=0.6]{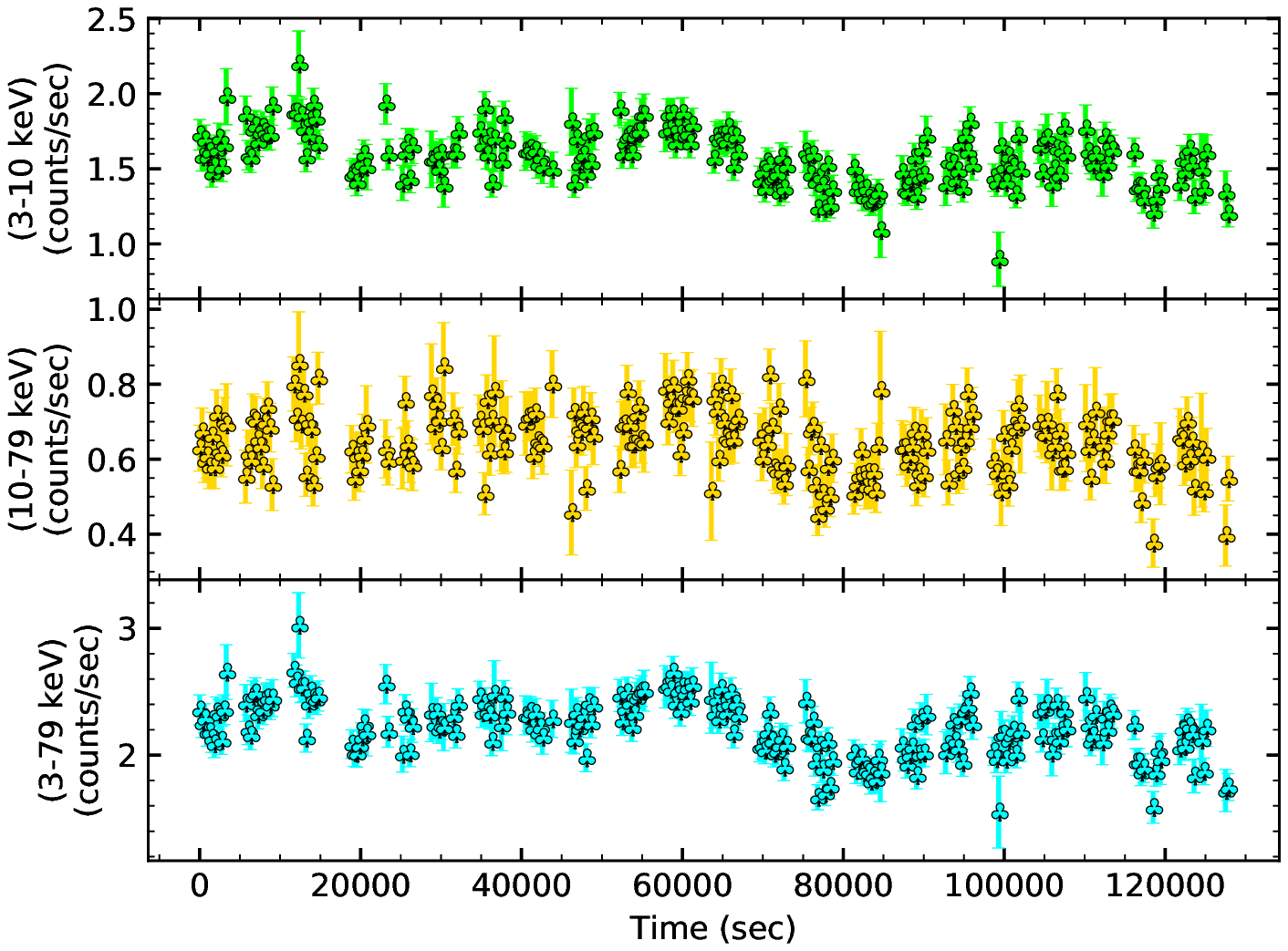}
            \hspace*{-0.5cm}\includegraphics[scale=0.6]{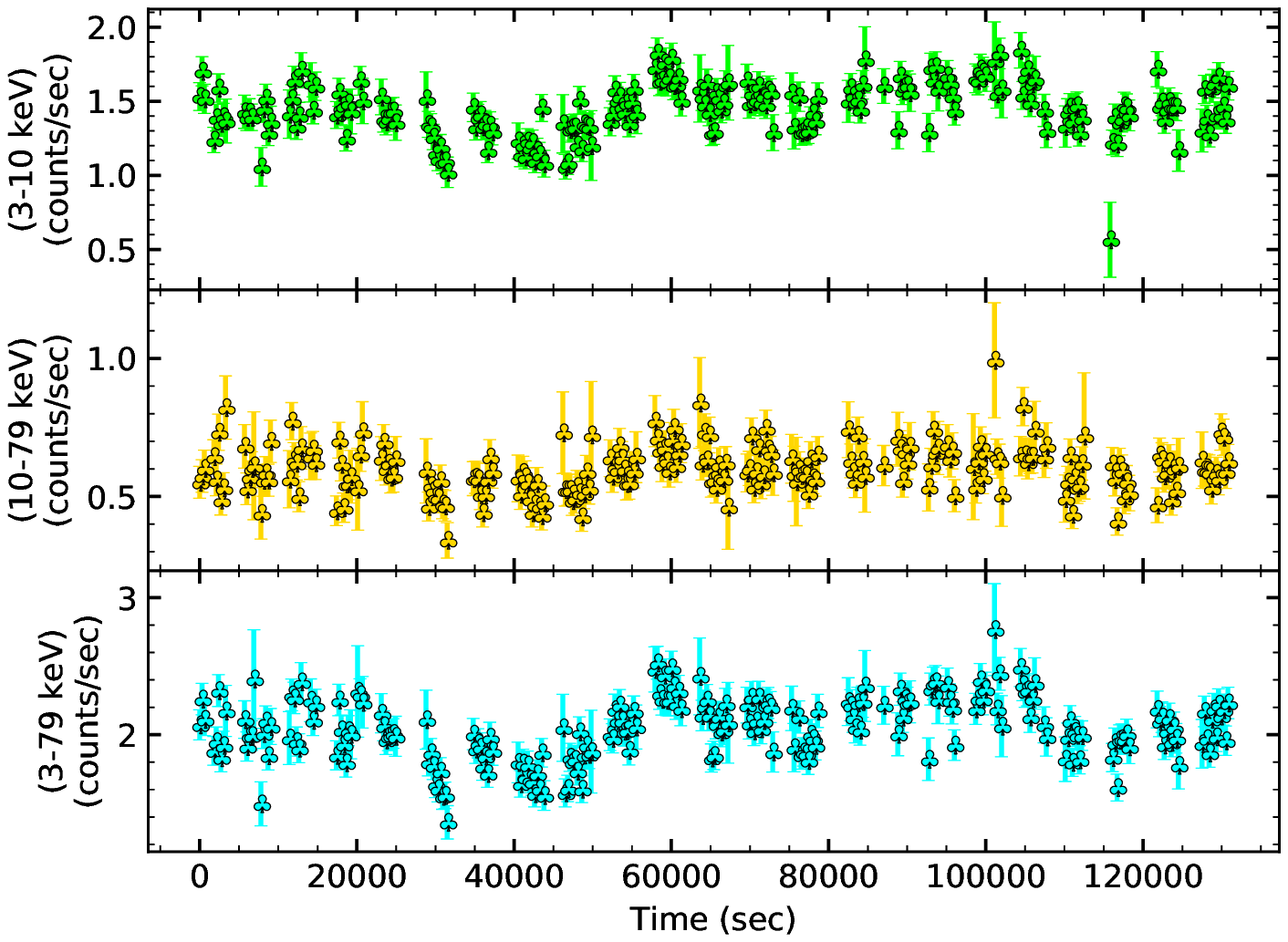}
            }
     \hbox{
           \hspace*{-0.5cm}\includegraphics[scale=0.6]{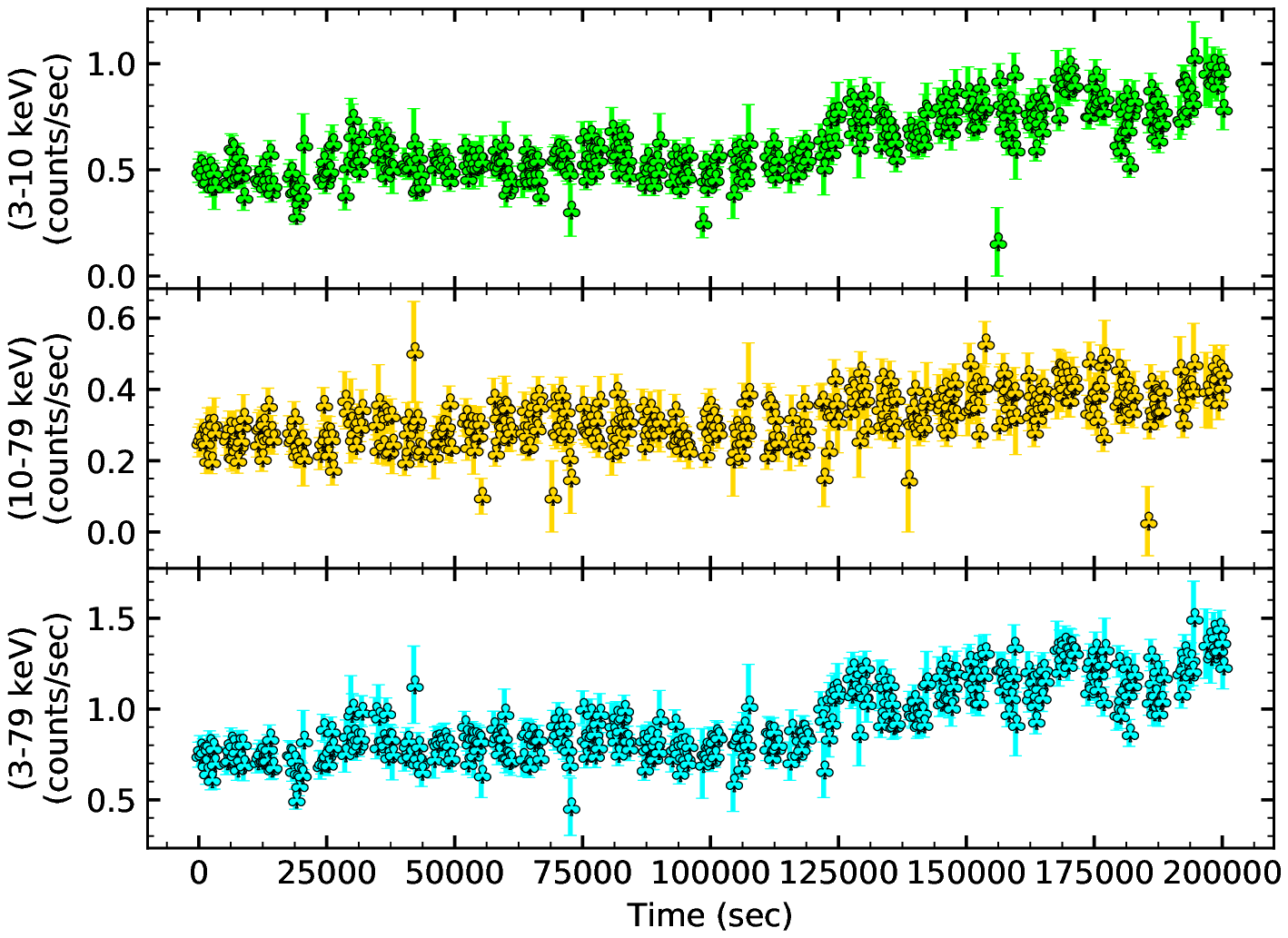}
           \hspace*{-0.5cm}\includegraphics[scale=0.6]{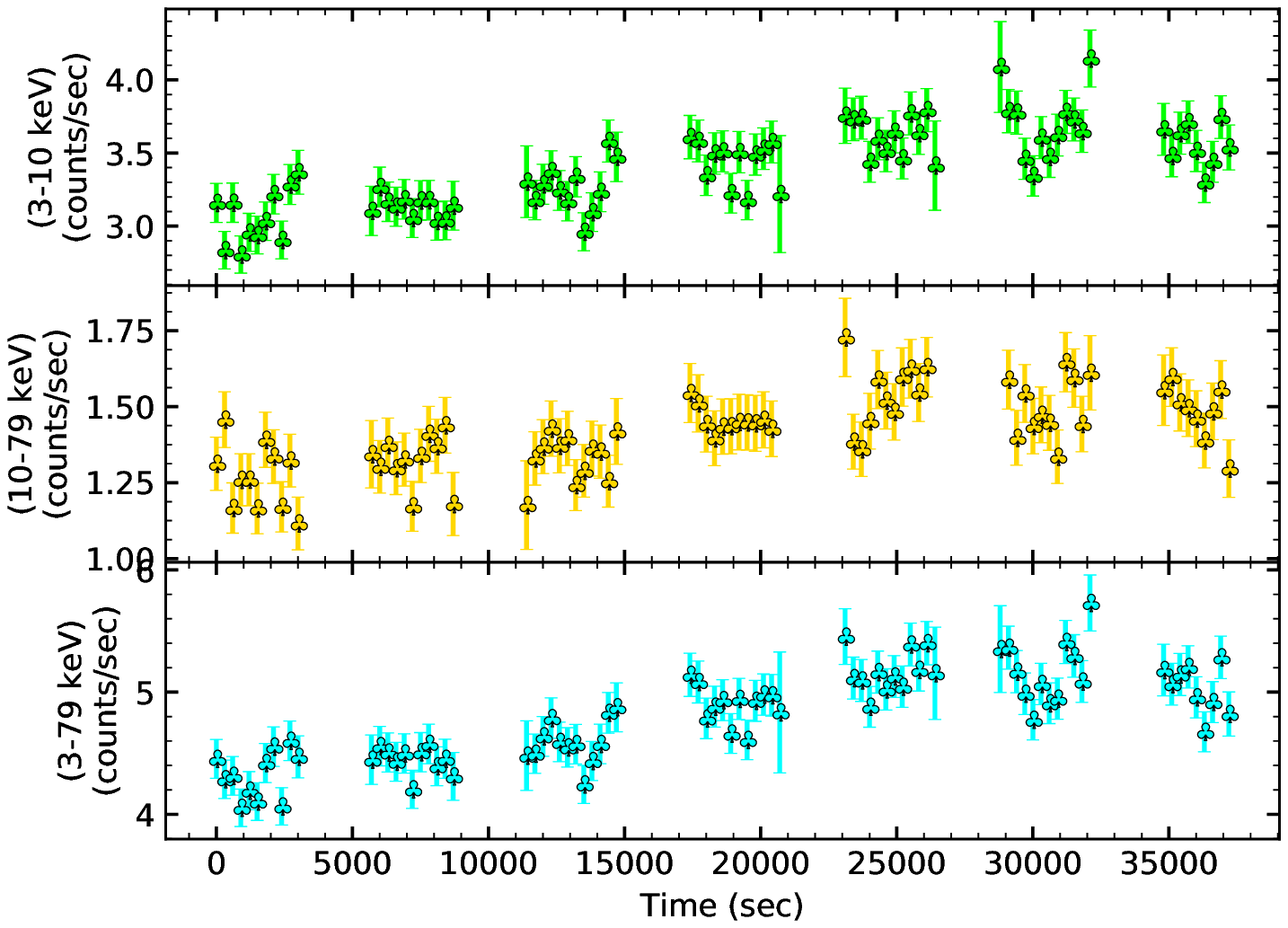}
          }
      \hbox{
      \hspace*{-0.5cm}\includegraphics[scale=0.6]{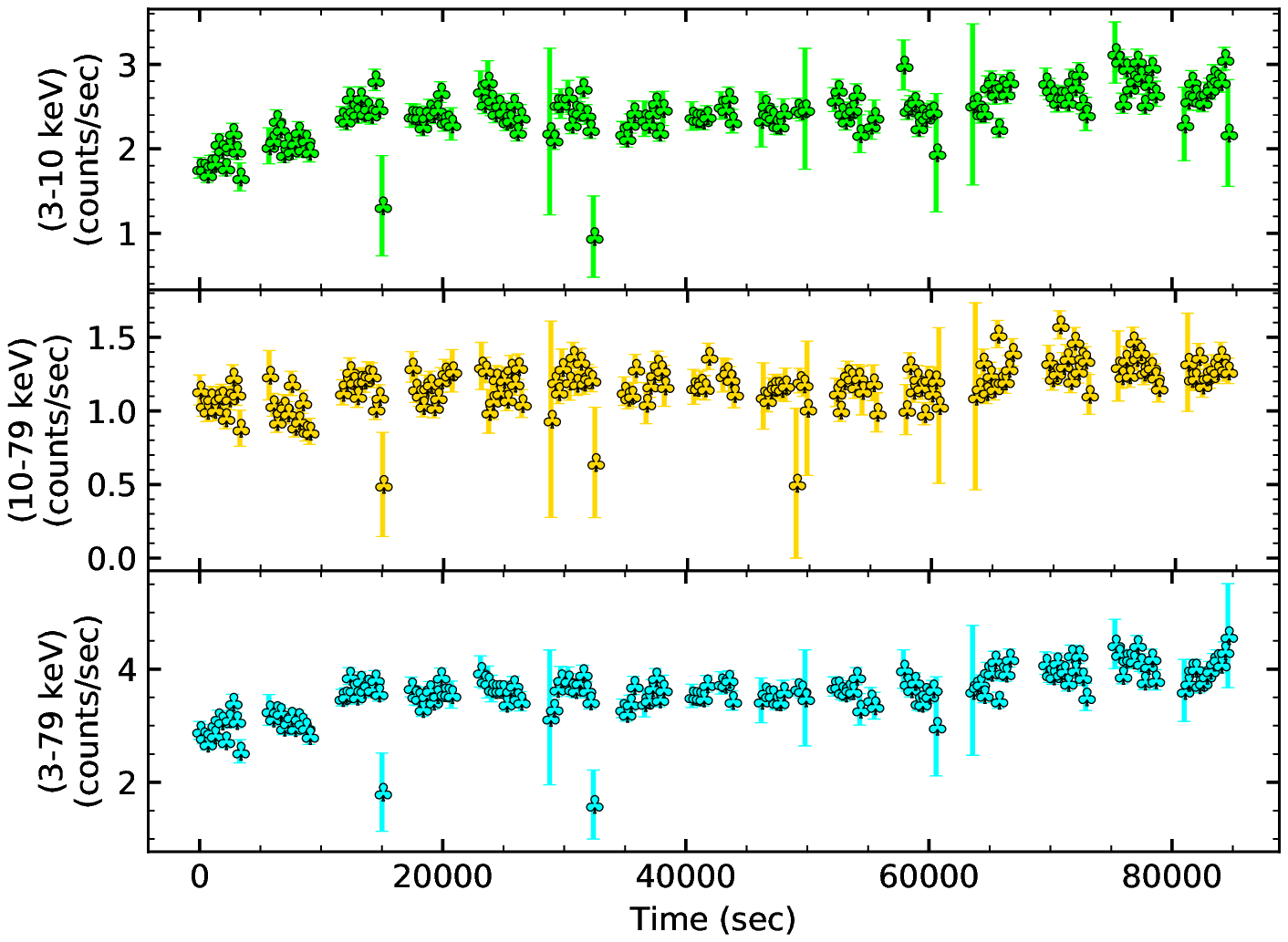}
      \hspace*{-0.5cm}\includegraphics[scale=0.6]{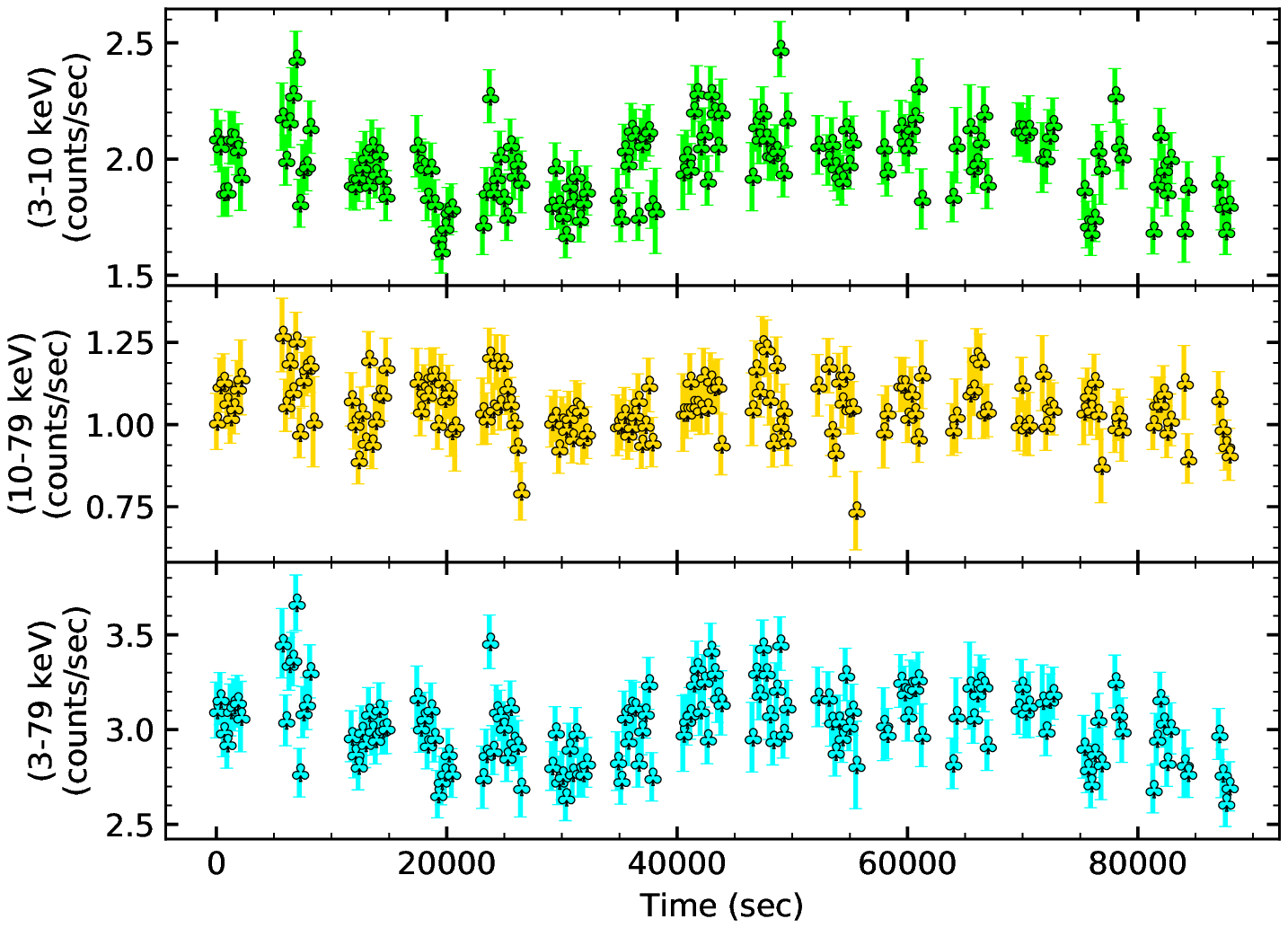}
      }
 }

\caption{\label{Fig-2} {\it NuSTAR} light curves in 3$-$10 keV, 10$-$79 keV and 3$-$79 keV bands. The top panel is for the source Mrk 1040 for the OBSID 60101002002 (left) and 
OBSID 60101002002 (right). The middle panel shows the light curves of the sources ESO 362$-$G18 (left) and NGC 2992 (right). The bottom panel shows the light curves
of the source NGC 3783 for OBSID 60101110002 (left panel) and 60101110004 (right panel).}
\end{figure*}

\begin{table*}
\caption{\label{log-HR}Results of linear least squares fit between HR and the count rate in the 3$-$79 keV band. Columns are
(1) name, (2) observational IDs, (3) slope and the error in slope (4) intercept and error in intercept (5) reduced $\chi^2$, 
(6) probability for no correlation and (7) linear correlation coefficient}
\centering
\begin{tabular}{@{}llrlllr@{}}
\hline 
Name            & OBSID         & Slope                & Intercept           & $\chi^2_{red}$ & P                     & R \\ 
 (1)            & (2)           & (3)                  & (4)                 & (5)       & (6)                        & (7) \\
\hline
Mrk 348         &  60160026002  &   0.000$\pm$0.028    &  0.565$\pm$0.113    &  1.086   &  0.96                        &  $-$0.005 \\
Mrk 1040        &  60101002002  &  $-$0.039$\pm$0.015    &  0.489$\pm$0.035    &  1.050   &  0.03                      &  $-$0.145 \\
                &  60101002002  &  $-$0.019$\pm$0.015    &  0.441$\pm$0.031    &  0.907   &  0.52                      &  $-$0.043 \\
ESO362$-$G18    &  60201046002  &  $-$0.150$\pm$0.022    &  0.646$\pm$0.023    &  0.980   &  1.20 $\times$ 10$^{-8}$   &  $-$0.285 \\
NGC 2992        &  60160371002  &  $-$0.004$\pm$0.010    &  0.435$\pm$0.047    &  0.927   &  0.68                      &  $-$0.047 \\
NGC 3783        &  60101110002  &  $-$0.048$\pm$0.011    &  0.658$\pm$0.040    &  1.205   &  0.20                      &  $-$0.099 \\
                &  60101110004  &  $-$0.147$\pm$0.024    &  0.976$\pm$0.072    &  1.168   &  0.00                      &  $-$0.236 \\
4U 1344$-$60    &  60201041002  &  $-$0.017$\pm$0.006    &  0.493$\pm$0.021    &  1.065   &  0.08                      &  $-$0.089 \\
ESO 141$-$G055  &  60201042002  &  $-$0.033$\pm$0.012    &  0.445$\pm$0.024    &  0.915   &  0.05                      &  $-$0.104 \\
Mrk 509         &  60101043002  &     0.025$\pm$0.008    &  0.305$\pm$0.028    &  1.089   &  0.65                      &   0.018 \\
                &  60101043004  &     0.074$\pm$0.018    &  0.163$\pm$0.057    &  1.218   &  0.06                      &   0.158 \\
NGC 7172        &  60061308002  &     0.022$\pm$0.010    &  0.442$\pm$0.047    &  0.874   &  0.06                      &   0.170 \\
NGC 7314        &  60201031002  &  $-$0.036$\pm$0.003    &  0.467$\pm$0.009    &  1.088   &  7.02 $\times$ 10$^{-28}$  &  $-$0.530 \\
\hline

\end{tabular}
\end{table*}

\begin{table*}
\caption{\label{log-model1}Best fitting model parameters for the sources using the model TBabs$\times$zTbabs$\times$pow. Columns are (1) name, (2) OBSID, 
(3) galactic column density in units of 10$^{20}$ cm$^{-2}$, (4) intrinsic column density in units of 10$^{22}$ cm$^{-2}$, 
(5) X-ray photon index, (6) normalization factor and (7) reduced $\chi^2$ }
\centering
\begin{tabular}{@{}lllllll@{}}
\hline 
Name            &        OBSID          & $N_{H(TBabs)}$  &  $N_{H(zTBabs)}$         & $\Gamma$       & $N_{pow} \times 10^3$    & $\chi^2/dof$  \\
 (1)            &   (2)                 & (3)             & (4)                      & (5)            & (6)                      & (7) \\
\hline
Mrk 348         &       60160026002     &   1.76$^{*}$    &  2.20$_{-0.72}^{+0.73}$  &  1.57$\pm$0.02 &  2.31$\pm$0.13           &  1.25   \\
Mrk 1040        &       60101002002     &   7.23$^{*}$    & 1.63$\pm$0.37            &  1.86$\pm$0.02 & 9.97$_{-0.46}^{+0.49}$   &  0.79   \\
                &       60101002004     &   4.11$^{*}$    & 1.28$_{-0.39}^{+0.40}$   &  1.85$\pm$0.02 & 8.97$_{-0.45}^{+0.48}$   &  0.80   \\
ESO 362$-$G18   &       60201046002     &   1.76$^{*}$    & 2.20$_{-0.72}^{+0.73}$   &  1.57$\pm$0.02 & 2.31$\pm$0.13            &  1.25   \\
NGC 2992        &       60160371002     &   5.26$^{*}$    & 2.90$\pm$0.42            &  1.90$\pm$0.02 & 2.44$_{-0.13}^{+0.14}$   &  0.67   \\
NGC 3783        &       60101110002     &   8.26$^{*}$    & 2.12$\pm$0.35            &  1.72$\pm$0.02 & 1.25$\pm$0.05            &  0.88   \\
                &       60101110004     &   4.11$^{*}$    & 3.32$_{-0.40}^{+0.41}$   &  1.68$\pm$0.02 & 9.89$_{-0.46}^{+0.48}$   &  0.90   \\
4U 1344$-$60    &       60201041002     &   1.07$^{*}$    & 2.12$\pm$0.24            &  1.81$\pm$0.01 & 1.44$\pm$0.04            &  1.08   \\
ESO141G055      &       60201042002     &   5.11$^{*}$    & 0.70$\pm$0.33            &  1.88$\pm$0.02 & 8.71$_{-0.35}^{+0.37}$   &  0.91   \\
Mrk 509         &       60101043002     &   4.11$^{*}$    & 1.08$\pm$0.27            &  1.81$\pm$0.01 & 1.41$\pm$0.03           &  1.35   \\
                &       60101043004     &   4.11$^{*}$    & 1.22$_{-0.61}^{+0.62}$   &  1.77$\pm$0.02  & 1.19$\pm$0.06           &  1.13   \\
NGC 7172        &       60061308002     &   1.65$^{*}$    & 9.95$\pm$0.44            &  1.83$\pm$0.02  & 2.34$_{-0.10}^{+0.11}$  &  0.82   \\
NGC 7314        &       60201031002     &   1.46$^{*}$    & 0.77$\pm$0.37            &  1.87$\pm$0.01  & 1.18$\pm$0.04           &  1.43   \\

\hline 

\end{tabular}
\end{table*}

\begin{table*}
\caption{\label{log-model2}Best fitting model parameters for the sources using the model TBabs$\times$zTbabs$\times$(zgauss+pexrav). However, for sources, Mrk 348, NGC 2992 and NGC 7172, {\it zgauss} is not 
used. The columns are: (1) Name of the sources, (2) OBSIDs, (3) peak of the Fe K$\alpha$ line in 
keV, (4) width of the Fe K$\alpha$ line in keV, (5) photon index, (6) $E_{cut}$ in keV, (7) reflection fraction, (8) normalization in units of 10$^{-2}$ and (9) $\chi^2$ per degree of freedom}
\centering
\begin{tabular}{@{}lllllllll@{}}
\hline 
Name        &   OBSID               &  $E$ (keV)              &  $\sigma$ (keV)         & $\Gamma$                 &   E$_{\mathrm{cut}}$ (keV)   &  $R$                     &   N$_{pexrav}$            &  $\chi^2/dof$ \\ 
(1)         & (2)                  & (3)                      & (4)                     & (5)                      & (6)                    & (7)                      &  (8)                      & (9) \\
\hline
Mrk 348         &       60160026002     &     ---             &   ---                  & 1.68$\pm$0.05           &   79$_{-19}^{+39}$   &  0.38$_{-0.22}^{+0.26}$  &   1.61$_{-0.10}^{+0.12}$  &  0.67 \\
Mrk 1040        &       60101002002     &  6.35$_{-0.05}^{+0.05}$ & 0.11$_{-0.11}^{+0.07}$ &  1.91$\pm$0.04          & 99$_{-22}^{+39}$     &  0.88$_{-0.23}^{+0.26}$  &   1.07$_{-0.05}^{+0.06}$  &  0.75 \\
                &       60101002004     &  6.44$_{-0.09}^{+0.10}$ & 0.30$_{-0.11}^{+0.13}$ &  1.94$\pm$0.04 & 114$_{-30}^{+61}$    &  0.95$_{-0.25}^{+0.29}$  &   0.99$_{-0.05}^{+0.06}$  &  0.76 \\
ESO 362$-$G18   &       60201046002     &  6.33$_{-0.04}^{+0.04}$ & 0.13$_{-0.07}^{+0.06}$ &  1.71$_{-0.05}^{+0.03}$ & $>$241               &  0.70$_{-0.14}^{+0.26}$  &   0.27$_{-0.02}^{+0.01}$  &  0.97 \\
NGC 2992        &       60160371002     &    ---              &  ---                       &  1.84$\pm$0.04          & 150$_{-65}^{+129}$   &  0.07$_{-0.07}^{+0.23}$  &   2.28$_{-0.12}^{+0.13}$  &  0.67 \\
NGC 3783        &       60101110002     &  6.24$_{-0.06}^{+0.05}$ &  0.12$_{-0.12}^{+0.08}$ & 1.88$\pm$0.04          & 77$_{-11}^{+15}$     &  1.86$_{-0.32}^{+0.37}$  &   1.52$_{-0.08}^{+0.09}$  &  0.79 \\
                &       60101110004     &  6.30$_{-0.04}^{+0.03}$ &  0.00$_{-0.00}^{+0.11}$ & 1.87$\pm$0.04 & 63$_{-8}^{+11}$      &  2.46$_{-0.34}^{+0.50}$  &   1.25$\pm$0.08           &  0.80 \\
4U 1344$-$60    &       60201041002     &  6.36$_{-0.04}^{+0.04}$ &  0.12$_{-0.12}^{+0.07}$ & 1.95$\pm$0.03          & 91$_{-10}^{+13}$     &  1.54$_{-0.19}^{+0.20}$  &   1.71$\pm$0.06           &  0.92 \\
ESO141G055      &       60201042002     &  6.31$_{-0.06}^{+0.07}$ &  0.08$_{-0.07}^{+0.12}$ & 1.94$\pm$0.04          & 69$_{-10}^{+14}$     &  1.20$_{-0.24}^{+0.27}$  &   0.94$_{-0.04}^{+0.05}$  &  0.86 \\
Mrk 509         &       60101043002     &  6.40$_{-0.04}^{+0.04}$ &  0.14$_{-0.07}^{+0.06}$ & 1.83$\pm$0.02          & 160$_{-23}^{+31}$    &  0.27$_{-0.04}^{+0.05}$  &   1.46$\pm$0.04           &  1.10 \\
                &       60101043004     &  6.40$_{-0.06}^{+0.09}$ &  0.00$_{-0.00}^{+1.09}$ & 1.78$\pm$0.04 & 143$_{-36}^{+72}$    &  0.23$_{-0.09}^{+0.10}$  &   1.20$_{-0.06}^{+0.07}$  &  1.06 \\
NGC 7172        &       60061308002     &  ---                    &   ---                   & 1.87$\pm$0.04          & 69$_{-10}^{+14}$     &  1.09$_{-0.23}^{+0.26}$  &   2.51$_{-0.14}^{+0.15}$  &  0.80 \\
NGC 7314        &       60201031002     &  6.36$_{-0.09}^{+0.08}$ &   0.50$_{-0.10}^{+0.14}$& 2.03$\pm$0.03          & $--$                 &  1.02$_{-0.16}^{+0.18}$  &   1.40$\pm$0.05           &  1.05 \\

\hline 

\end{tabular}
\end{table*}

\begin{table*}
\caption{\label{log-full}Up to date list of sources having $E_{cut}$ measurements from {\it NuSTAR} and associated details. For sources that are analysed in 
this work and having more than one OBSID, the lowest values of $E_{cut}$ is given in the table. The values of $E_{cut}$,$\Gamma$, M$_{BH}$ and
$\lambda_{Edd}$ quoted in this table are taken from the references given in the last column. }
\centering
\begin{tabular}{@{}lllrllllllll@{}}
\hline 
No.& Name         & $\alpha_{2000}$ & $\delta_{2000}$  & $z$    & V        & Type   & $E_{cut}$           & $\Gamma$              & M$_{BH}$ & $\lambda_{Edd}$  & Reference \\  
   &              &                 &                  &        & (mag)    &        &  (keV)              &                       &                 &           & \\ \hline
1 &  Mrk 348           & 00:48:47.2   &    31:57:25.0    & 0.014  & 14.59    & Sy1h   & $79^{39}_{19}$      & 1.68 $\pm$ 0.05        & 7.2 & 0.149 & This work   \\ 
2 & Mrk 1040           & 02:28:14.4   &    31:18:41.0    & 0.016  & 14.74    & Sy1    & $99^{+39}_{-22}$    & 1.91 $\pm$ 0.04        & 6.4 & 1.030 & This work   \\ 
3   & 3C 120           & 04:33:11.1   &    05:21:15.0    & 0.033  & 15.05    & Sy1.5  & $83^{+10}_{-08}$    & 1.87 $\pm$0.02         & 7.7 & 0.353 & A           \\
4   & Ark 120          & 05:16:11.4   & $-$00:09:00.0    & 0.033  & 13.92    & Sy1    & $183^{+83}_{-43}$   & 1.87 $\pm$ 0.02        & 8.2 & 0.085 & C,J         \\
5   & ESO 362$-$G18    & 05:19:35.8   & $-$32:39:27.0    & 0.013  & 13.37    & Sy1.5  & $>$ 241             & $1.71^{+0.03}_{-0.05}$ & 7.7 & 0.012 & This work   \\
6   & MCG +8-11-11     & 05:54:53.6   &    46:26:21.0    & 0.020  & 14.62    & Sy1.5  & $175^{+110}_{-50}$  & 1.77 $\pm$ 0.04        & 7.2 & 0.754 & C,H         \\
7   & NGC 2992         & 09:45:42.0   & $-$14:19:35.0    & 0.008  & 13.78    & Sy1.9  & $150^{+129}_{-65}$  & 1.84 $\pm$ 0.04        & 7.7 & 0.029 & This work   \\ 
8     & MCG-5-23-16    & 09:47:40.2   & $-$30:56:54.0    & 0.008  & 13.69    & Syi    & $116^{+6}_{-5}$     & 1.85 $\pm$ 0.01        & 7.8 & 0.031 & A           \\
9    & NGC 3783        & 11:39:01.8   & $-$37:44:19.0    & 0.009  & 13.43    & Sy1.5  & $63^{+11}_{-8}$     & 1.87 $\pm$ 0.04         & 6.9 & 0.146 & This work   \\  
10   & NGC 4151        & 12:10:32.5   &    39:24:21.0    & 0.003  & 11.85    & Sy1.5  & 59 $\pm$ 4.0       & 1.66 $\pm$ 0.02         & 7.6 & 0.100 & B,K           \\
11   & PG 1247+268     & 12:50:05.7   &    26:31:07.0    & 2.042  & 15.92    & QSO    & $89^{+112}_{-34}$   & $2.35^{+0.09}_{-0.08}$ & 8.9 & 0.024 & C,I           \\
12   & NGC 5273        & 13:42:08.3   &    35:39:15.0    & 0.003  & 13.12    & Sy1.9  & $143^{-96}_{40}$    & $1.81^{+0.02}_{-0.03}$ & 6.8 & 1.10  & A           \\ 
13   & 4U 1344$-$60    & 13:47:36.0   & $-$60:37:03.0    & 0.013  & 19.00    & Sy1    & $91^{+13}_{-10}$    & 1.95 $\pm$ 0.03        & 8.2 & 0.014 & This work   \\ 
14   & IC 4329A        & 13:49:19.3   & $-$30:18:34.0    & 0.016  & 13.66    & Sy1.2  & 186 $\pm$ 14        & 1.73 $\pm$ 0.01        & 6.8 & 0.082 & A           \\
15   & NGC 5506        & 14:13:14.8   & $-$03:12:26.0    & 0.007  & 14.38    & Sy1i   & $720^{+130}_{-190}$ & 1.91 $\pm$ 0.03        & 8.0 & 0.013 & A           \\
16   & GRS 1734-292    & 17:37:28.3   & $-$29:08:02      & 0.021  & 21.0     & Sy1    & $53^{+11}_{-08}$    & 1.65 $\pm$ 0.05        & 8.5 & 0.033 & A           \\
17   & 3C 382          & 18:35:03.4   &    32:41:47.0    & 0.058  & 15.39    & Sy1    & $214^{147}_{-63}$   & $1.68^{+0.03}_{0.02}$  & 9.2 & 0.109 & A           \\
18   & ESO 103-035     & 18:38:20.5   & $-$65:25:39.0    & 0.013  & 14.53    & Sy2    & $183^{+83}_{-43}$   & 1.87 $\pm$ 0.02        & 8.2 & 0.085 & D,G         \\
19   & 3C 390.3        & 18:42:09.0   &    79:46:17.0    & 0.056  & 15.38    & Sy1.5  & $117^{+18}_{14}$    & 1.70 $\pm$ 0.01        & 8.4 & 0.240 & A           \\
20   & ESO141-G55      & 19:21:14.3   & $-$58:40:13.0    & 0.037  & 13.64    & Sy1.2  & $69^{+14}_{-10}$    & 1.94 $\pm$ 0.04        & 7.5 & 0.370 & This work   \\
21   & NGC 6814        & 19:42:40.7   & $-$10:19:23      & 0.005  & 14.21    & Sy1.5  & $155^{+70}_{-35}$   & $1.71^{+0.04}_{-0.03}$ & 7.0 & 0.003 & C,H         \\
22   & 4C 74.26        & 20:42:37.3   &    75:08:02.0    & 0.104  & 15.13    & Sy1    & $183^{+51}_{-35}$   & $1.84^{+0.03}_{-0.02}$ & 9.6 & 0.037 & A           \\
23   & Mrk 509         & 20:44:09.7   & $-$10:43:24.0    & 0.035  & 13.12    & Sy1.5  & $143^{+72}_{-36}$   & 1.78 $\pm$ 0.04        & 7.9 & 0.215 & This work   \\ 
24   & IGR 2124.7+5058 & 21:24:39.4   &    50:58:25.0    & 0.020  & 15.4 R   & Sy1    & $80^{+11}_{-09}$    & 1.59 $\pm$ 0.02        & 7.5 & 0.400 & E,G           \\
25   & J2127.4+5654    & 21:27:44.9   &    56:56:40      & 0.014  & 18.79    & Sy1n   & $108^{+11}_{-10}$   & 2.08 $\pm$ 0.01        & 7.2 & 0.090 & A           \\
26   & NGC 7172        & 22:02:01.9   & $-$31:52:08.0    & 0.009  & 13.61    & Sy2    & $69^{+14}_{-10}$    & 1.87 $\pm$ 0.04        & 8.3 & 0.004 & This work   \\
27   & QSO B2202-209   & 22:05:09.9   & $-$01:55:18.0    & 1.770  & 17.50    & QSO    & $153^{+103}_{-54}$  & 1.82 $\pm$ 0.05        & 9.1 & 1.150  & A           \\
28   & NGC 7314        & 22:35:46.1   & $-$26:03:02.0    & 0.005  & 13.11    & Sy1h   &                     & 2.03 $\pm$ 0.003       & 5.9 & 0.181 & This work   \\
29   & Ark 564         & 22:42:39.3   &    29:43:32.0    & 0.025  & 14.16    & S3     & 42 $\pm$ 3          & 2.27 $\pm$ 0.08        & 6.4 & 1.100  & A           \\
30   & NGC 7469        & 23:03:15.6   &    08:42:26.0    & 0.017  & 13.04    & Sy1.5  & $170^{+60}_{-40}$   & 1.78 $\pm$ 0.02        & 7.0 & 0.300  & F           \\
\hline 
\end{tabular}
\noindent A:\cite{2018ApJ...856..120R}; B: \cite{2018JApA...39...15R}, C: \cite{2018A&A...614A..37T}; D: \cite{2009MNRAS.392.1124V}, 
E:\cite{2010ApJ...721.1340T}, F: \cite{2018A&A...615A.163M} G: \cite{2018MNRAS.481.4419B}, H:\cite{2018A&A...614A..37T}, I:\cite{2016A&A...590A..77L},
J:\cite{2018A&A...609A..42P},K:\cite{2002ApJ...579..530W} \\
\end{table*}

%%%%%--------------------Main body-------------------------%%%%% 
%----------------------------1---------------------------------

\section{Sample, Observation and reduction}
Our sample of objects for the present study were selected from  the archives of 
the High Energy Astrophysics
Archive Research Center (HEASARC) \footnote{https://heasarc.gsfc.nasa.gov/cgi-bin/W3Browse/w3browse.pl}. 
We looked at the HEASARC archives for observations from {\it NuSTAR} that are open for use between
the period June 2013 - June 2018. From this, we initially focussed on 10 nearby objects, 
that  are also reasonably bright with net count rate in the 3$-$79 keV band 
greater than 0.1. The details of these 10 objects selected for this study are
given in Table \ref{log}.  

We reduced the data using {\it NuSTAR} Data Analysis Software package {\it NuSTARDAS} 
v1.6.0\footnote{https://heasarc.gsfc.nasa.gov/docs/nustar/analysis/nustar\_swguide.pdf} and CALDB version 20161207 distributed by  
HEASARC. We generated cleaned and screened event files using the 
{\it nupipeline} task, and also considering the passage of the satellite through 
the South Atlantic Anomaly(SAA). We extracted the spectra and corresponding 
response files using {\it nuproducts} task, with a circular region 
of $60''$ at the peak of the source and $60''$ radius circular background region away 
from the source on the same chip. 

For spectral analysis, we fitted both the focal plane modules FPMA and FPMB spectra simultaneously allowing the cross 
normalization for both modules to vary. In this fitting process, the abundances of the elements 
were fixed to their solar values \citep{1989GeCoA..53..197A}. 
We used {\it XSPEC} (version 12.9.0; \citealt{1996ASPC..101...17A}) for 
the spectral fitting. The $\chi^{2}$ minimization technique 
in {\it XSPEC} was used to get the best model description of the data and all errors were 
calculated using $\chi^{2}$ = 2.71 criterion i.e. 90\% confidence range for one 
parameter of interest.

For timing analysis, we generated light curves in the energy ranges of 
3$-$10 keV (soft-band), 10$-$79 keV (hard band) and 3$-$79 keV (total band) in both 
the focal plane modules using a time bin of 300 seconds. The light curves from the two
modules were them combined using the task {\it lcmath} available in 
FTOOLS V6.19.

\begin{figure*}
\vbox{

     \hbox{
           \hspace*{-0.5cm}\includegraphics[scale=0.6]{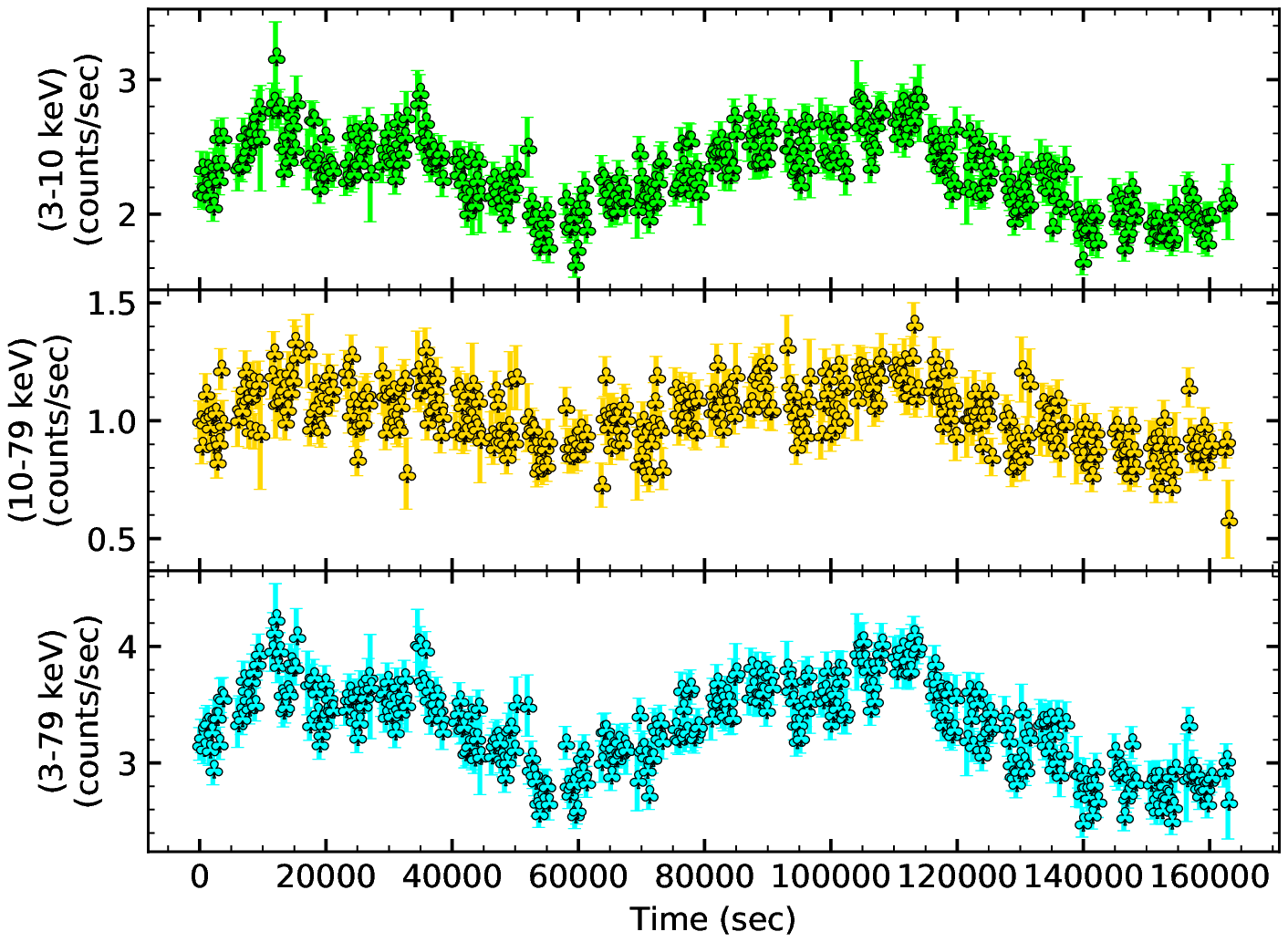}
           \hspace*{-0.5cm}\includegraphics[scale=0.6]{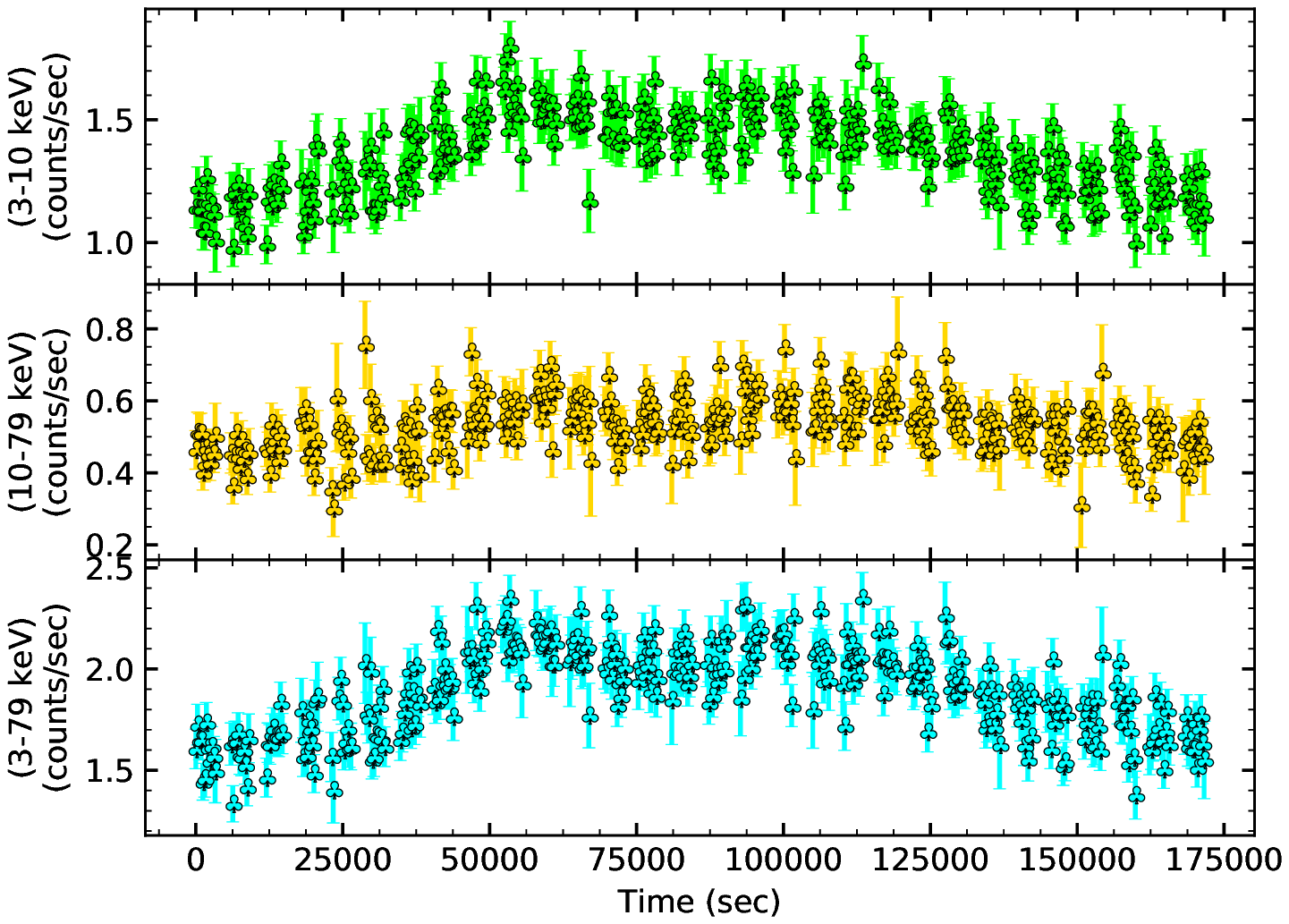}
          }
     \hbox{
     \hspace*{-0.5cm}\includegraphics[scale=0.6]{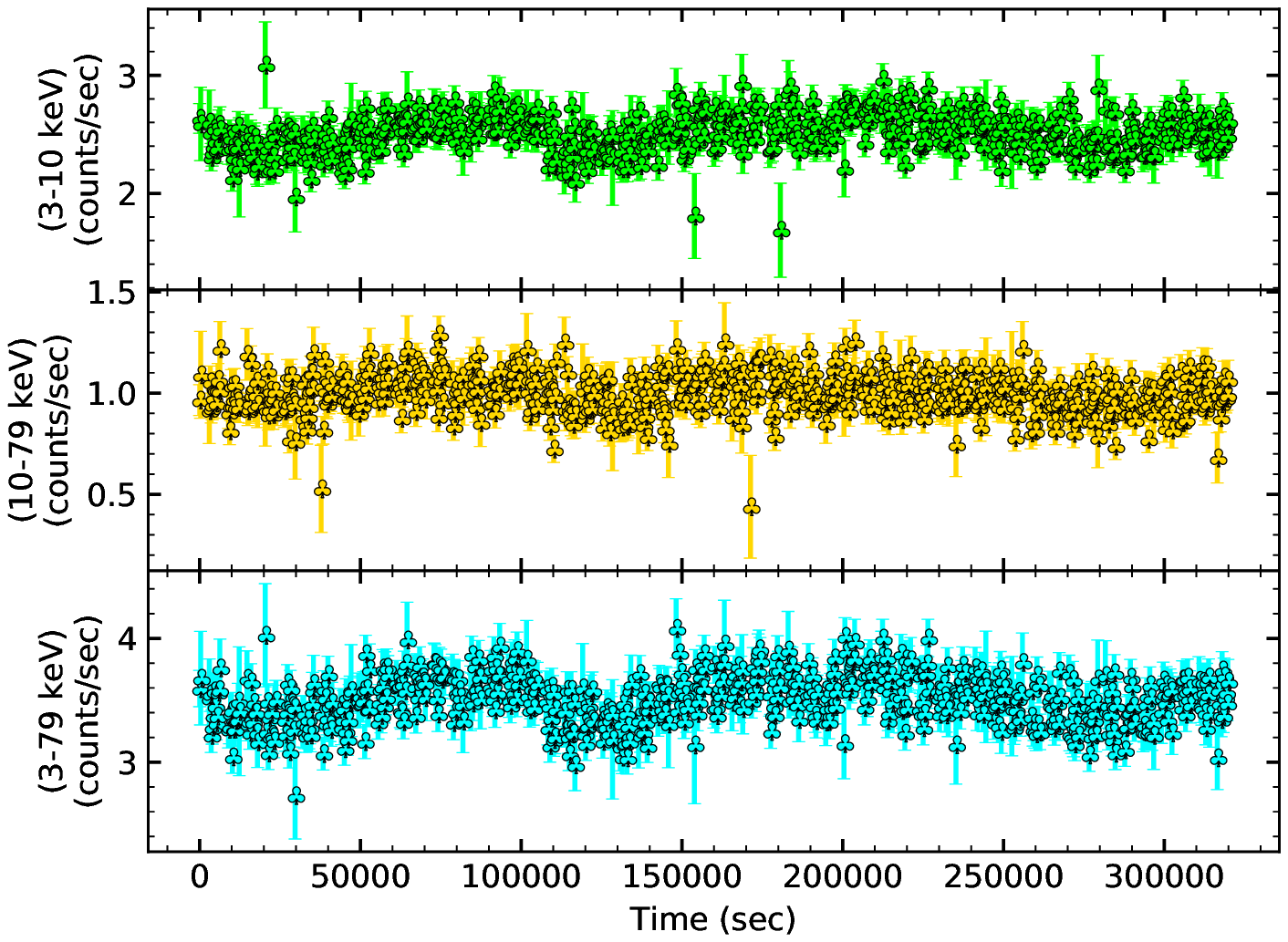}
     \hspace*{-0.5cm}\includegraphics[scale=0.6]{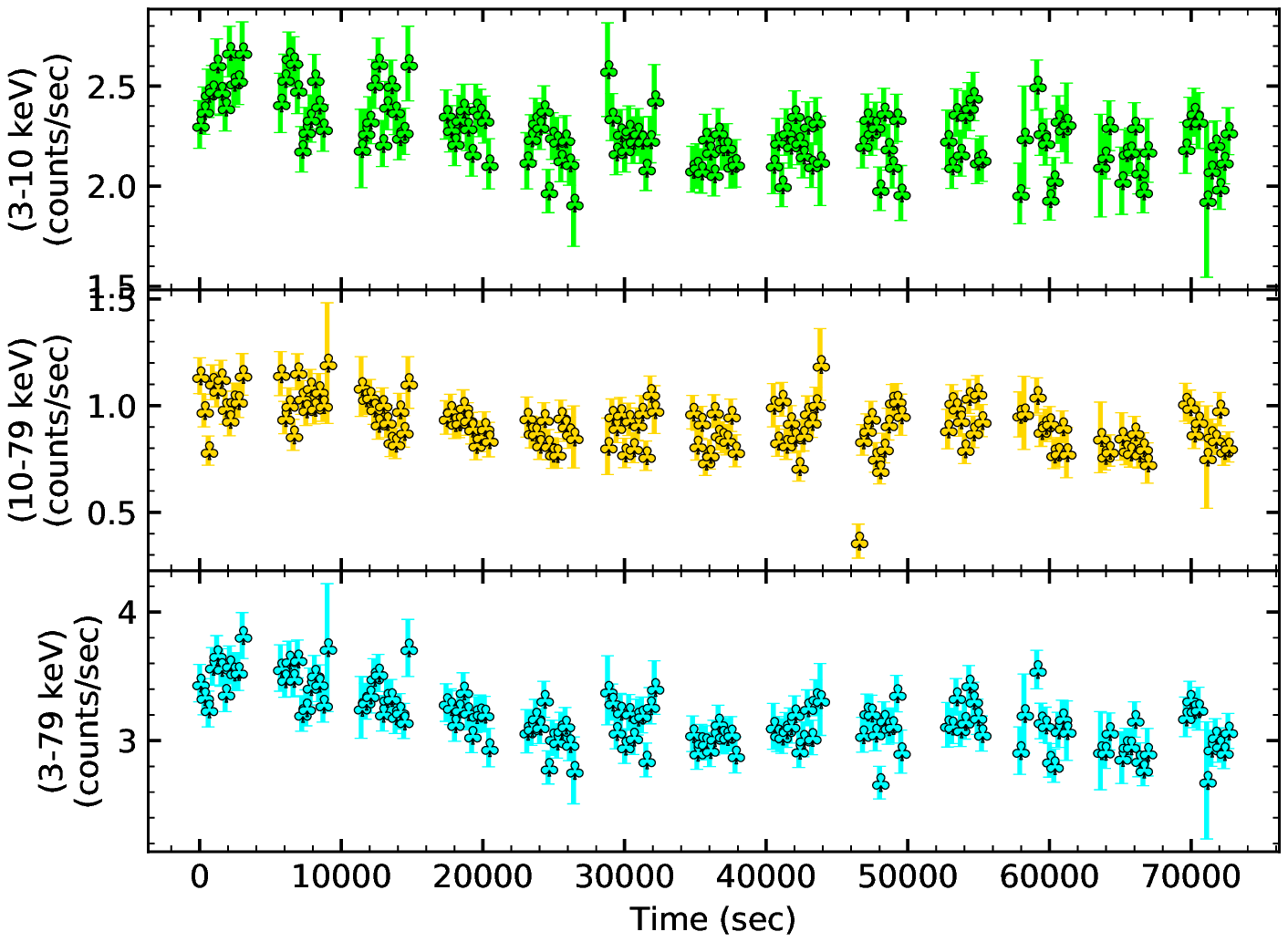}
        }
    \hbox{
    \hspace*{-0.5cm}\includegraphics[scale=0.6]{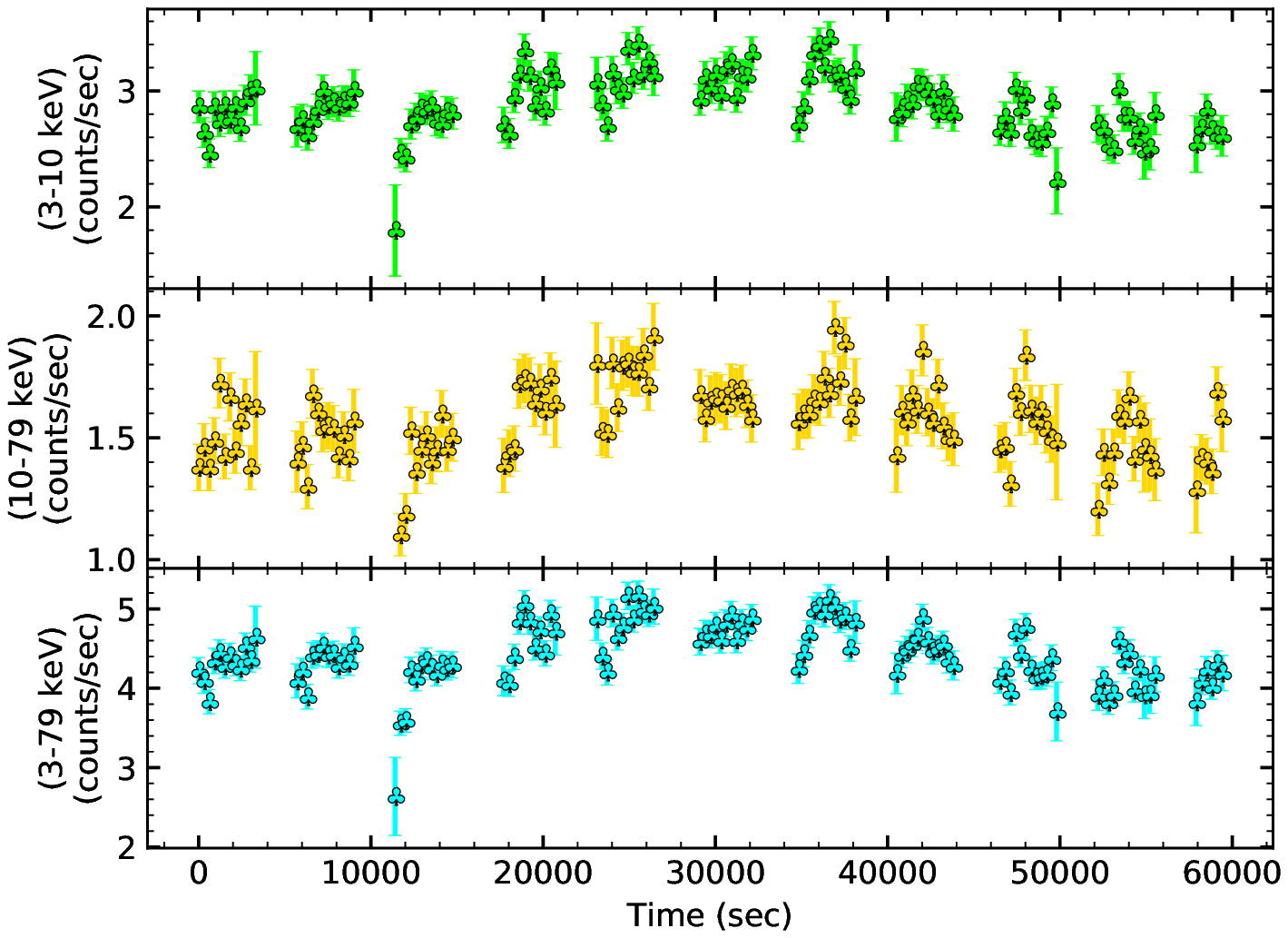}
    \hspace*{-0.5cm}\includegraphics[scale=0.6]{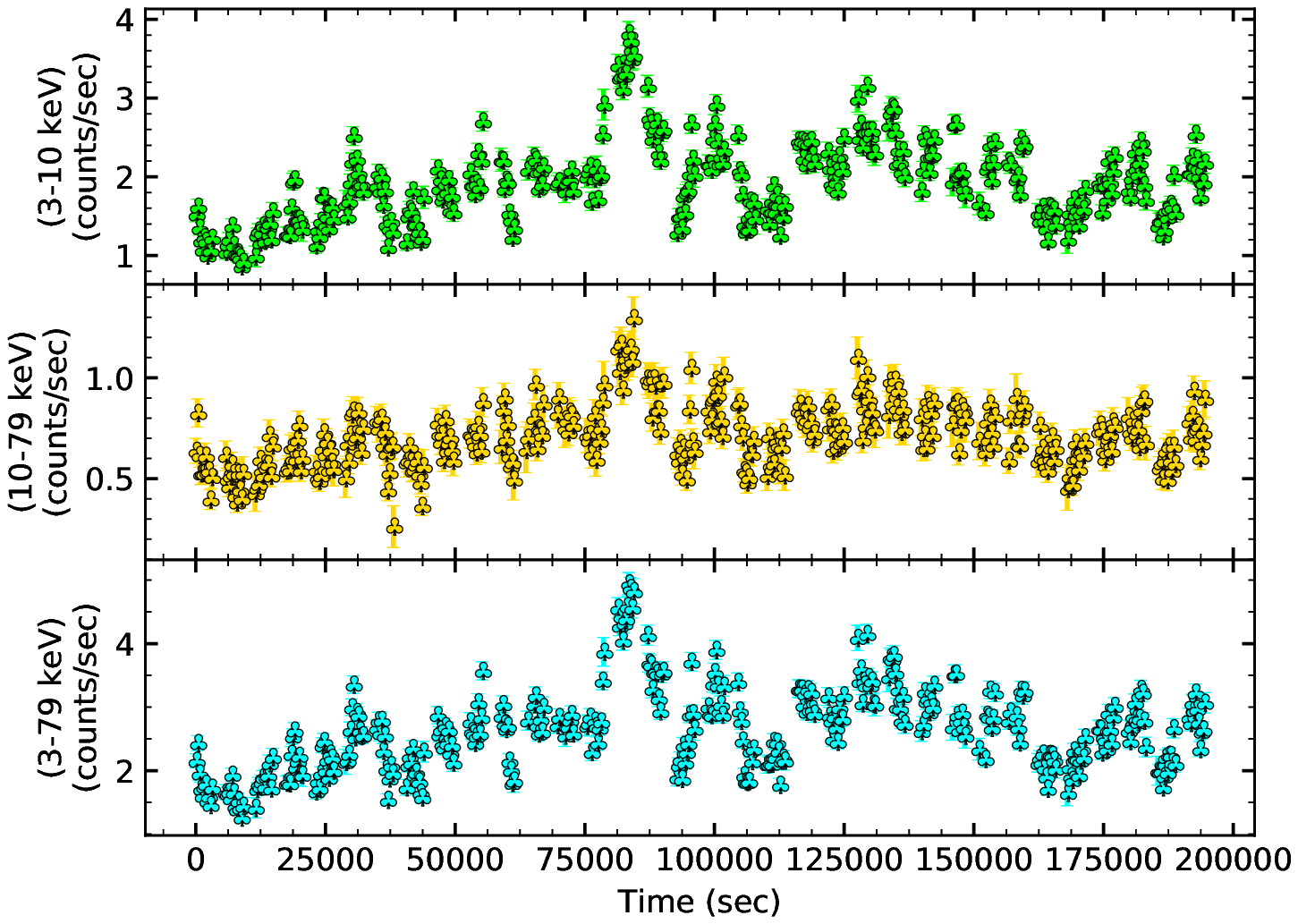}
         }
     }
\caption{\label{Fig-3}Light curves in 3$-$10 keV, 10$-$79 keV and 3$-$79 keV from {\it NuSTAR}. The top panel shows the light curves for the sources
4U 1344$-$60 (left) and ESO 141G055 (right). The middle panel shows the light curves for Mrk 509 for the OBSID 60101043002 (left) and the OBSID 60101043004 (right). The bottom
panel gives the light curves for the sources NGC 7172 (left) and NGC 7314 (right).}
\end{figure*}

\section{Results}
\subsection{Timing analysis}
For timing analysis the light curves generated by the procedures outlined in Section 2
was used. They were visually inspected to see any outlier points and for points with 
very large error bars. To remove both outliers and those with large error bars, for further
light curve analysis, the mean and standard deviation of the light curve as well as the 
mean and standard deviation of the errors were calculated. For each object, 
only those points satisfying the following two conditions were retained for further analysis, namely,
(a) the error on each individual data points should be  less than 5 times the standard deviation of 
the errors (b) the difference between a data point and the mean of the light curve should be less than 
5 times the standard deviation of the light curve. The final light curves of the 10 objects
studied here are given in Figures \ref{Fig-1} $-$ \ref{Fig-3}. Among the 10 objects, for three objects namely
Mrk 1040, NGC 3783 and Mrk 509, we have two sets of observations (OBSIDs) each, while 
for the remaining 6 objects we have one OBSID each. Visual inspection of the light curves
shown in Figures \ref{Fig-1} $-$ \ref{Fig-3} clearly show the objects to be variable. To characterise 
variability we calculated the normalized excess variance ($F_{var}$). $F_{var}$
\citep{2002ApJ...568..610E,2003MNRAS.345.1271V}, gives the intrinsic variation of the 
source after removal of the measurement errors. 
Following \cite{2003MNRAS.345.1271V} we define $F_{var}$ as
\begin{equation}
\centering
F_{\rm{var}}=\sqrt{{S^{2}-{\bar{\sigma^{2}}_{\rm{err}}}\over\bar{x}^2}}
\end{equation}
where $S^2$ represents the sample variance, $\bar{x}$ is the arithmetic mean of $x_i$
and $\bar{\sigma^{2}}_{\rm{err}}$ represents the mean square error.
$S^2$ and ${\bar{\sigma^{2}}_{\rm{err}}}$ are given as
\begin{equation}
\centering
S^2={1\over{N-1}}\sum_{i=1}^{N}(x_i-\bar{x})^2
\end{equation}
\begin{equation}
\bar{\sigma^{2}}_{\rm{err}}={1\over{N}}\sum_{i=1}^{N}\sigma^{2}_{\rm{err,i}}
\end{equation}
The uncertainty in $F_{\rm{var}}$ is calculated as
\begin{equation}
 \centering
          err(F_{\rm{var}})=\sqrt{\Bigg(\sqrt{1\over{2N}}{\bar{{\sigma^{2}}_{\rm{err}}}\over{{\bar{x}}^2}{F_{\rm{var}}}}\Bigg)^2+\Bigg(\sqrt{\bar{\sigma^{2}}_{\rm{err}}\over{N}}{1\over\bar{x}}\Bigg)^2}
\end{equation}
The results of the variability analysis is
given in Table \ref{log-variability}. From Table \ref{log-variability} it is evident that all sources showed variability in the soft, hard and total bands.
The flux variations in the soft band is found to be larger than that of the soft band in six sources. For three objects
namely Mrk 348, NGC 7172, and NGC 2992,  within error bars the flux variations in the soft and hard
bands are similar. For Mrk 509, flux variations in the hard band is larger than the soft band in both the
sets of observations. Considering all the sources together, the weighted mean variability amplitude 
in the soft, hard and total bands are 0.097 $\pm$ 0.081, 0.077 $\pm$ 0.050 and 0.090 $\pm$ 0.064 respectively. Thus
within error bars, the mean variability in all the bands are identical.  Also, statistical analysis for
similarities or differences in the F$_{var}$ between soft and hard bands for the whole sample was carried out using the 
non-parametric Mann-Whitney U test. The null hypothesis that was tested in this U-test is that 
the distribution of $F_{var}$ values of soft and hard bands are identical. At any given confidence level, the null hypothesis
is rejected if U is less than the critical U-value ($U_{crit}$).  For our sample, we found a U value of 65 and a 
critical U value of 45. At the $\alpha$ = 0.05 significance, as U is greater than $U_{crit}$, the null hypothesis is not rejected.
Thus, considering all the objects as a whole, there is no difference in the flux variability behaviour between 
soft and hard bands, but  when sources are considered individually difference in the flux variations between soft and hard bands is noticed.

\begin{figure*}
\hspace*{-0.5cm}\includegraphics{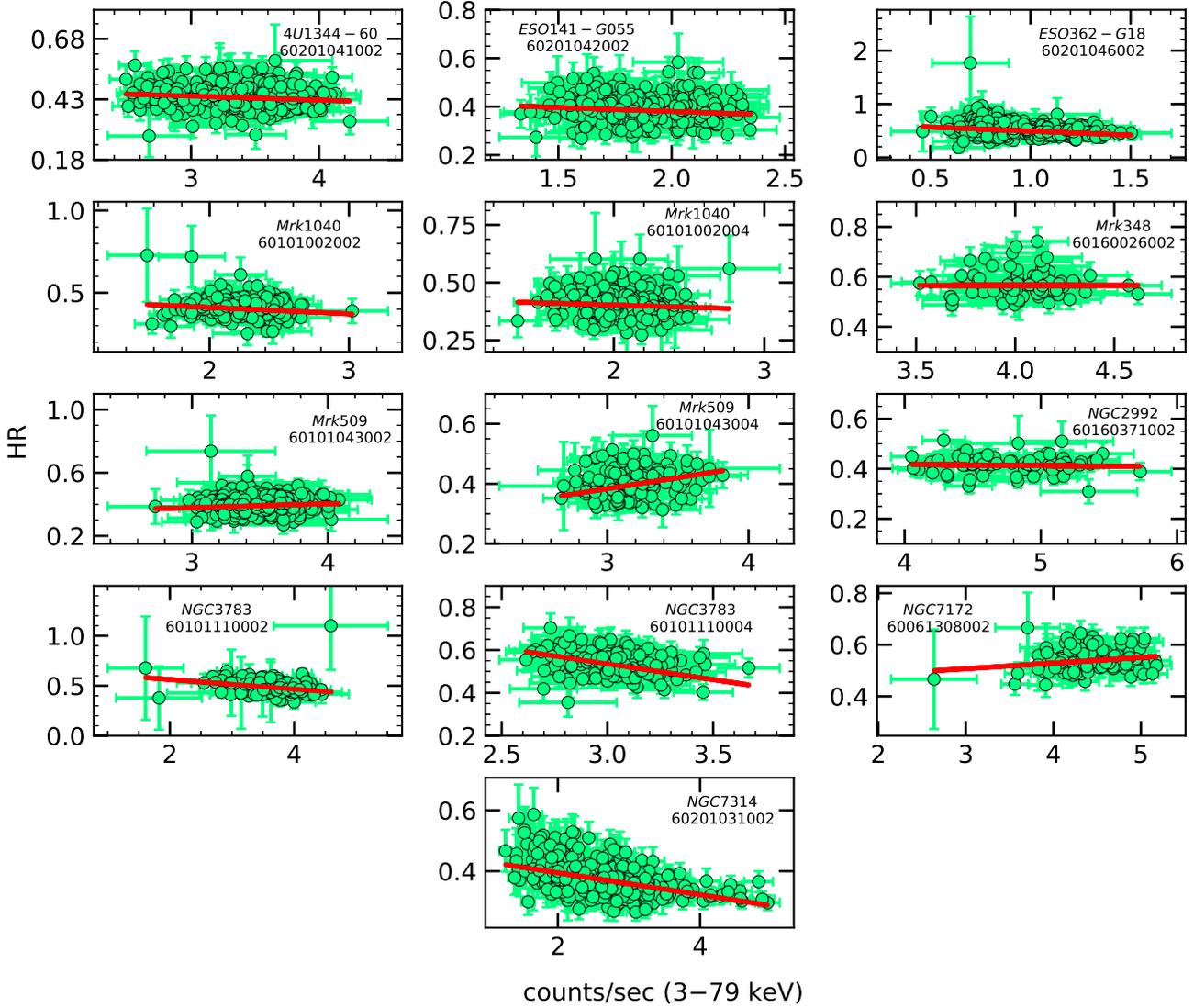}
\caption{\label{ratio}Plot of HR against the count rate in the 3$-$ 79 keV band. The red solid line is the weighed linear least squares fit to the data.}
\end{figure*}

To check for spectral variability in any, we calculated the hardness radio (HR)  and plotted it against the count rate in the total energy band
of 3$-$79 keV. We define HR as  
HR  = $F_{hard}/F_{soft}$ where, $F_{hard}$ and $F_{soft}$ are the count rates in the hard and soft bands respectively.
The plots of HR v/s the total count rate are shown in Fig. \ref{ratio}. Over plotted on the observations are linear least squares fit the to
the data using $HR = a  \times F_{3-79 keV} + b$ taking into account the errors in both HR and count-rate. There are indications of spectral 
variability, with most of the sources showing very weak but insignificant softer when brighter trend.  Two sources namely Mrk 509 and NGC 7172 
show very weak harder when brighter trend, but again it is insignificant. The only source that showed significant spectral variation is NGC 7314. Linear
least squares fit to the data of NGC 7314 shows a mild softer when brighter trend following the relation HR = ($-$0.036 $\pm$ 0.003) $\times$ F$_{3-79 keV}$ + 
(0.467 $\pm$ 0.009) with a linear correlation coefficient of $-$0.53. The softer when brighter trend inferred from timing analysis
is also known from spectral studies of AGN, wherein the X-ray spectra of Seyfert galaxies are found to be softer when they are brighter \citep{2016MNRAS.463..382U}.

\subsection{Spectral Analysis}
Our motivation in this work is to increase the number of AGN with $E_{cut}$ measurements and to check for correlations if any between
$E_{cut}$ values and other physical properties of the sources. Therefore, we employed two phenomenological model fits to the data.

\begin{figure}
  \includegraphics[scale=0.4]{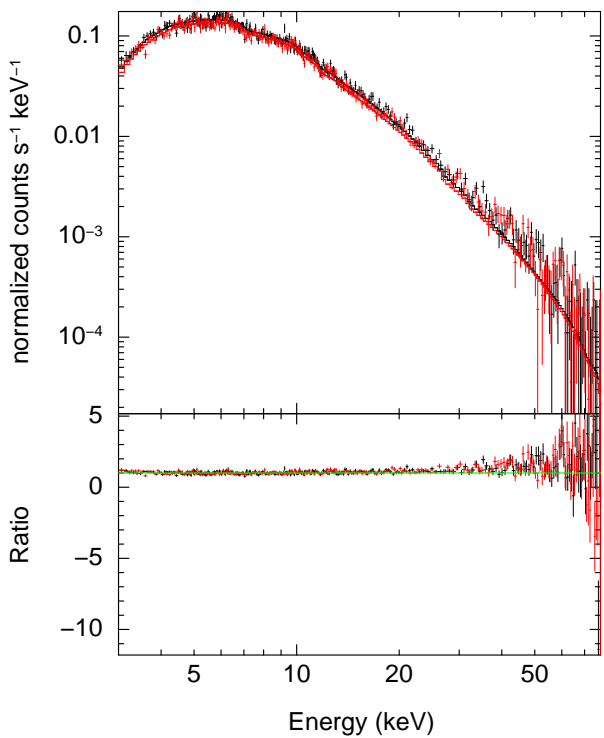}
  \caption{\label{Spec-1} Normalized counts/sec versus energy for the model {\tt TBabs$\times$z Tbabs$\times$(pexrav)} given for both FPMA (black)
and FPMB (red) modules for the source Mrk 348.The ratio plot in the bottom panel  gives the ratio of data to model for  
FPMA (black) and FPMB (red).}
\end{figure}

\begin{figure*}
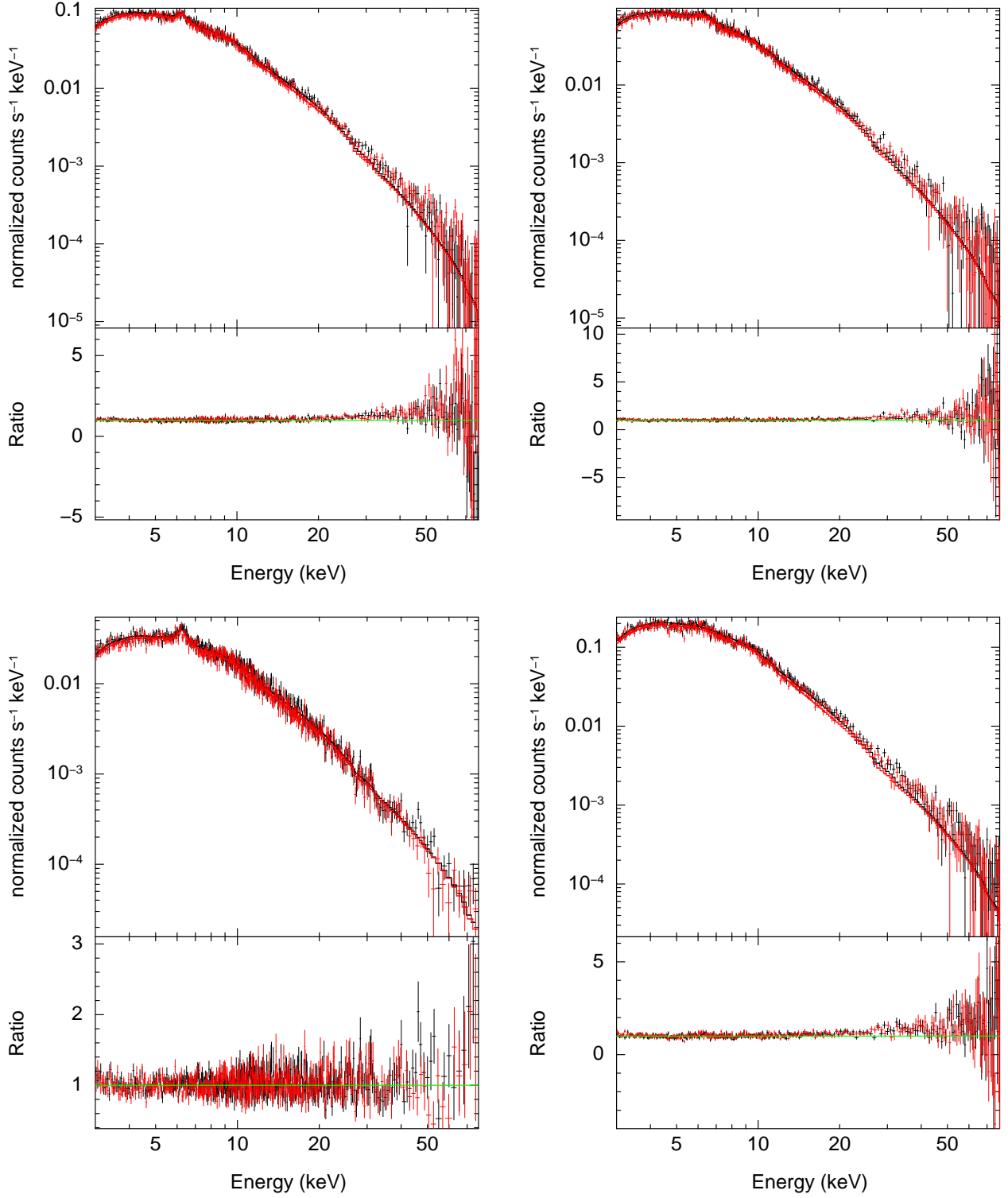

\vbox{
     \hbox {
           \includegraphics[scale=0.4]{spectra/mrk1040_02/model2.eps}
           \includegraphics[scale=0.4]{spectra/mrk1040_04/model2.eps}
            }
     \hbox{
         \includegraphics[scale=0.4]{spectra/eso362/model2.eps}
         \includegraphics[scale=0.4]{spectra/ngc2992/model2.eps}
         }
     }
\caption{\label{Spec-2}Observed spectra along with model fits TBabs$\times$zTbabs$\times$(zgauss+pexrav) and the ratio spectrum. The top panel is for the source Mrk 1040 for the OBSID 60101002002 (left) and the 
OBSID 60101002004 (right). The bottom panel is for the sources ESO 362$-$G18 (left) and NGC 2992 (right). For NGC 2992 the fitted model is TBabs$\times$zTbabs$\times$(pexrav)}
\end{figure*}

\begin{figure*}
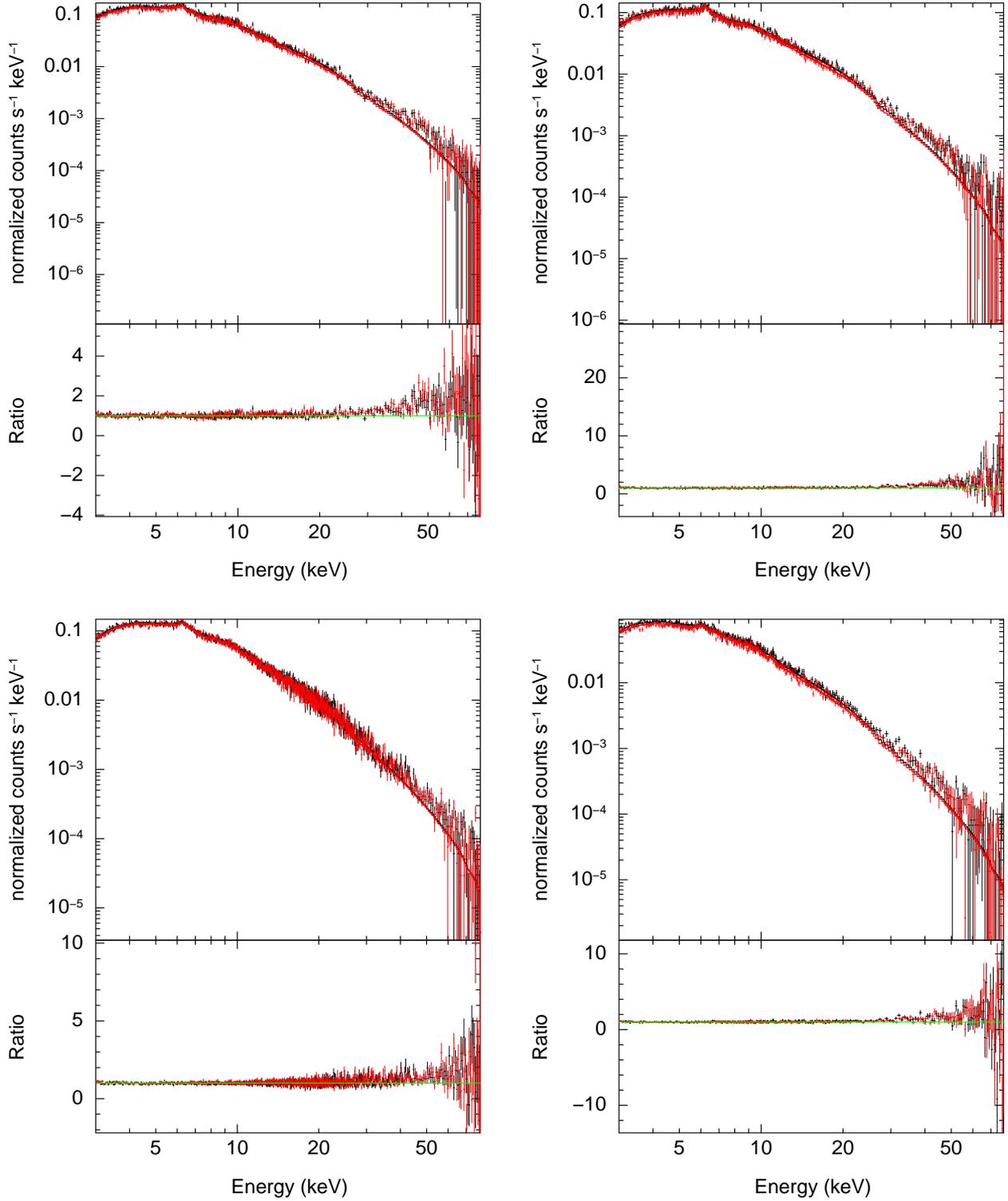

\vbox{
      \hbox{
           \includegraphics[scale=0.4]{spectra/ngc3783_02/model2.eps}
           \includegraphics[scale=0.4]{spectra/ngc3783_04/model2.eps}
            }
      \hbox{
           \includegraphics[scale=0.4]{spectra/4u_1344m60/model2.eps}
           \includegraphics[scale=0.4]{spectra/eso_141mg_55/model2.eps}
            }
     }
 \caption{\label{Spec-3}Normalized counts/sec versus energy for the model {\tt TBabs$\times$zTbabs$\times$(zgauss+pexrav)} given for both FPMA (black) 
and FPMB (red) modules and the ratio plots. The top panel is for the source NGC 3783 for the OBISD 60101110002 (left) and 60101110004(right). The bottom
panel is for the sources 4U 1344$-$60 (left) and ESO 141G055 (right).}
\end{figure*}

\begin{figure*}
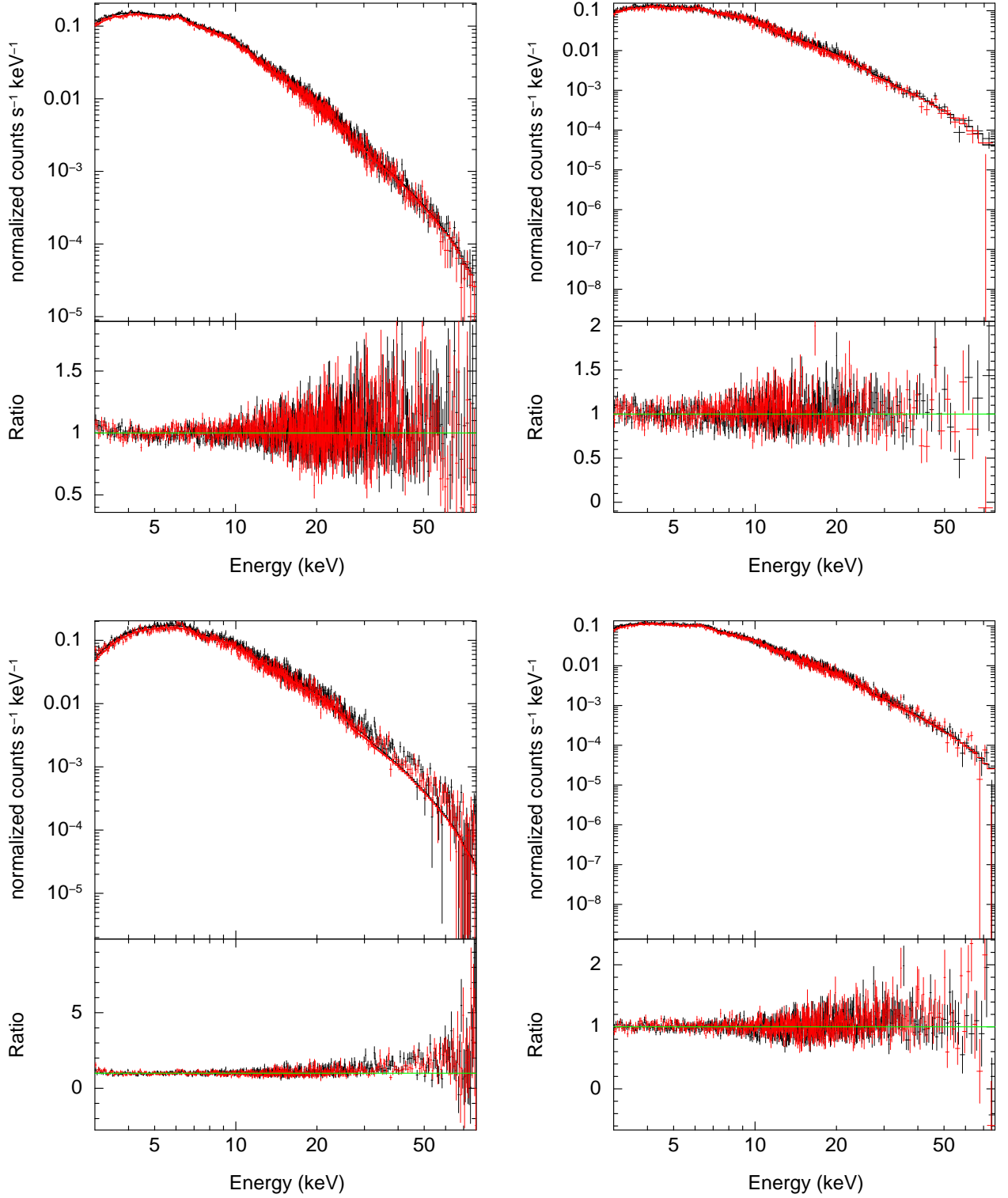

\vbox{
       \hbox{
            \includegraphics[scale=0.4]{spectra/mrk509_02/model2.eps}
            \includegraphics[scale=0.4]{spectra/mrk509_04/model2.eps}
            }
       \hbox{
            \includegraphics[scale=0.4]{spectra/ngc7172/model2.eps}
            \includegraphics[scale=0.4]{spectra/ngc7314/model2.eps}
            }
   }
  \caption{\label{Spec-4}Observed {\it NuSTAR} spectra along with the model fit using the model {\tt TBabs$\times$zTbabs$\times$(zgauss+pexrav)} given for both FPMA (black) 
and FPMB (red) modules and the ratio spectra.  The top panel is for the source Mrk 509 for the OBSID 60101043002 (left) and the OBSID 60101043004 (right).
The bottom panel is for the source NGC 7172 (left) and NGC 7314 (right). For the source NGC 7172 the {\tt zgauss} component of the model was not used.}
\end{figure*}

\subsubsection{Model-1}
We first used the simple absorbed power law model {\tt TBabs$\times$zTBabs$\times$powlaw} to fit each of the AGN spectra. 
{\tt TBabs} \citep{2000ApJ...542..914W} was used to model the Galactic absorption whereas {\tt zTBabs} 
was used to consider the absorption due to host galaxy of the source. For this model, we 
used \cite{1989GeCoA..53..197A} set of solar abundances and the \cite{1992ApJ...400..699B} photoelectric cross 
sections. The galactic neutral hydrogen column density was frozen to the value  
obtained from \cite{1990ARA&A..28..215D} for all the sources. In this model the free parameters were the photon index $\Gamma$ and
the normalization. 
In some of the sources, we found evidence of iron K$\alpha$ line and reflection component in the residuals, along with 
high energy turnover.  
The  fitting results along with the galactic neutral hydrogen column density that was used and frozen during the fit  are given in Table \ref{log-model1}.

\subsubsection{Model-2}
We noticed turnover in the residuals obtained by fitting the model
{\tt TBabs$\times$zTBabs$\times$powlaw} to the data. This  clearly suggested of the presence of cut-off 
in the spectrum. 
Also, in the residual spectra of simple power law model (model-1) fits to the data there
were indications of the presence of the fluorescent Fe K$\alpha$ line. This line
is present in the X-ray spectra of most of the AGN \citep{1993ARA&A..31..717M},
consisting of both broad and narrow components. Therefore, Fe K$\alpha$  component was
included in the spectral analysis of the sources analysed here. From model-1 fits, we found
that for three sources namely Mrk 348, NGC 2992 and NGC 7172, the Fe K$\alpha$ line was not
visibly in their observed spectra. Therefore, for those three sources, while fitting model-2, 
the Gaussian component to model the Fe K$\alpha$ line was not used, while it was used
in the other 7 sources. The parameters of the component
that were extracted from the spectral analysis are the peak
energy of the line, the width of the line and the normalization.
Also, in the observed hard X-ray emission of AGN, both $E_{cut}$ and reflection are believed to 
play an important role. Therefore to obtain $E_{cut}$, we replaced the   {\tt powerlaw} in model-1 with the  {\tt Pexrav} component and refitted  
each AGN spectra.  {\tt Pexrav}
\citep{1995MNRAS.273..837M} includes both primary emission in the form of a power law with an exponential cut-off and
the reflection component,  wherein it calculates the spectrum of the X-ray source on reflection from an optically thick neutral slab. In this model, the output parameter $R$, gives a measure of the 
reflection component present in the observed spectrum. If the source is isotropic, $R$ is related to the 
solid angle as $R \sim \Omega/2 \pi$ and this value of $R$ depends on the angle of inclination $i$ between the 
perpendicular to the accretion disk and the line of sight to the observer. 
During the spectral fitting, we used the default value of the inclination angle of $i$ = 45$^\circ$ 
and abundances present in the model. As $i$ is fixed to the default value for all the
fitting, the values of $R$ derived from the fit only gives an indication of the amplitude of reflection. The nH values for the {\tt zTBabs} component of the model was frozen to the value obtained from model-1.
The components that were left free during this model fit were $E_{cut}$, peak of the Fe K$\alpha$ line, standard deviation of the Fe K$\alpha$ line,
reflection parameter and normalization for both {\tt zguass} and {\tt pexrav} components of the model. 
The model fit along with the residual spectrum are shown in 
Figure \ref{Spec-1} $-$ \ref{Spec-4} and fitting results are given in Table \ref{log-model2}. 

In three out of the ten sources analysed here, namely, Mrk 348, NGC 2992 and 
NGC 7172 Fe K$\alpha$ line is not seen. In the standard model of AGN, broad 
Fe K$\alpha$ line is expected to be ubiquitously present in spectra of AGN, 
however, there are exceptions \citep{2011MNRAS.416..629B}. The apparent 
non-detection of Fe K$\alpha$ 
line in the spectra of AGN could be due to them viewed at large angles to 
the line of sight to the observer subsequently leading to weaker reflection 
\citep{2011MNRAS.416..629B}, low signal-to-noise ratio (S/N) spectra, very 
high ionised accretion disk \citep{1993MNRAS.261...74R,1994MNRAS.266..653Z} 
or a combination of the above. All the three sources for which Fe K$\alpha$ 
line is not seen here are viewed at larger angles having classification of 
Sy1h, Sy1.9 and Sy2 in the \cite{2010A&A...518A..10V} catalog respectively. Thus, the apparent 
lack of Fe K$\alpha$ line in them could be due to weaker reflection owing to 
larger viewing angle, however, more detailed spectral analysis is needed to 
clearly pin point the causes for the absence of Fe K$\alpha$ line in these 
sources. As the aim of this work is to find $E_{cut}$, detailed spectral 
analysis of the sources are not attempted here.

%%%%%%%%%%%%%%%%%%%%%%%%%%%%%%%%%%%%%%%%%%%%%%%%%%%%%%%%%%%%%%%%%
\begin{figure}
\hspace*{-0.5cm}\includegraphics[scale=0.6]{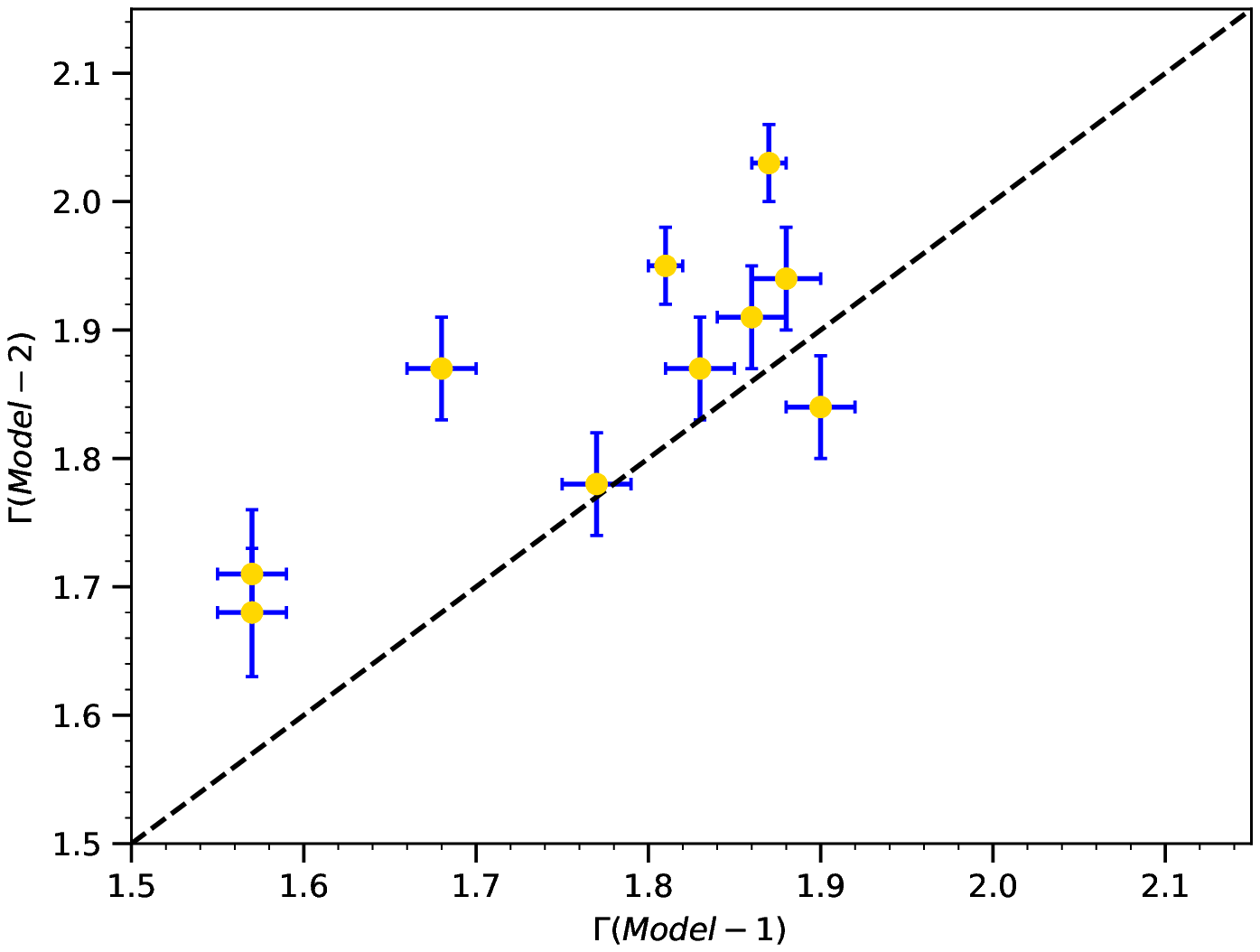}
\caption{\label{index} $\Gamma$ obtained from model-1 against $\Gamma$ obtained from model-2}
\end{figure}

\begin{figure}
\vbox{
\vspace*{-0.5cm}
\hspace*{-0.6cm}\includegraphics[scale=0.6]{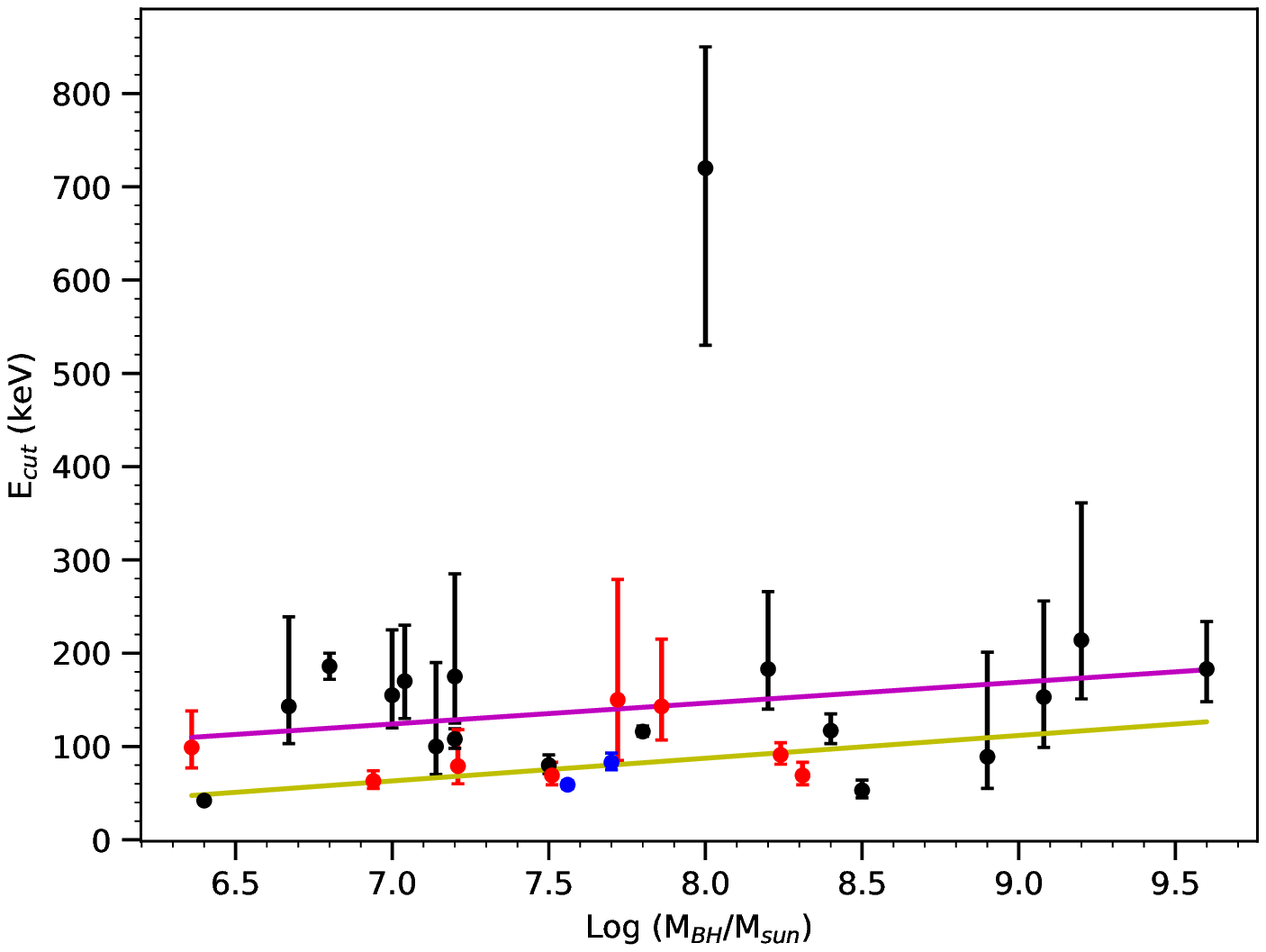}
\hspace*{-0.6cm}\includegraphics[scale=0.6]{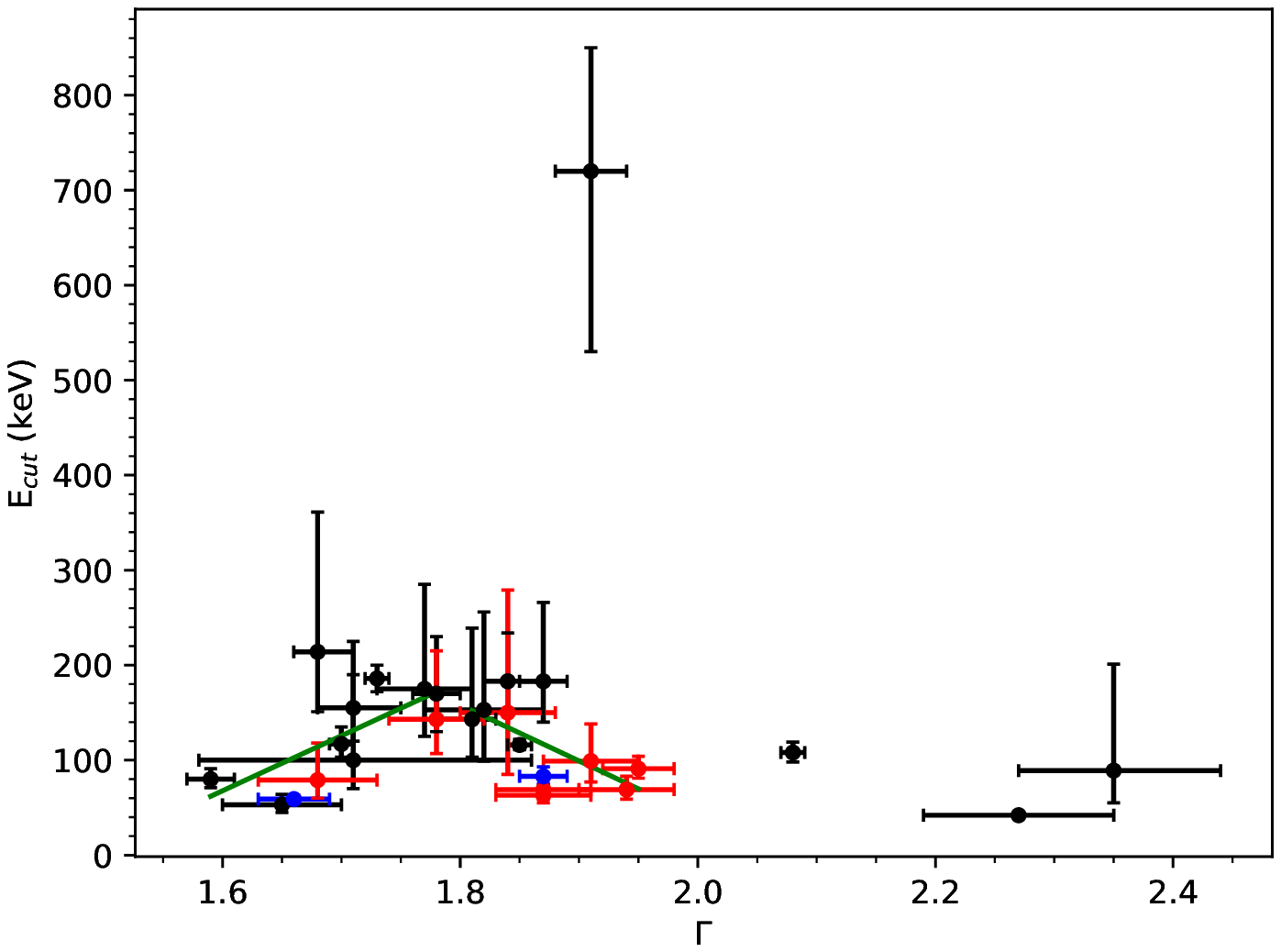}
\hspace*{-0.6cm}\includegraphics[scale=0.6]{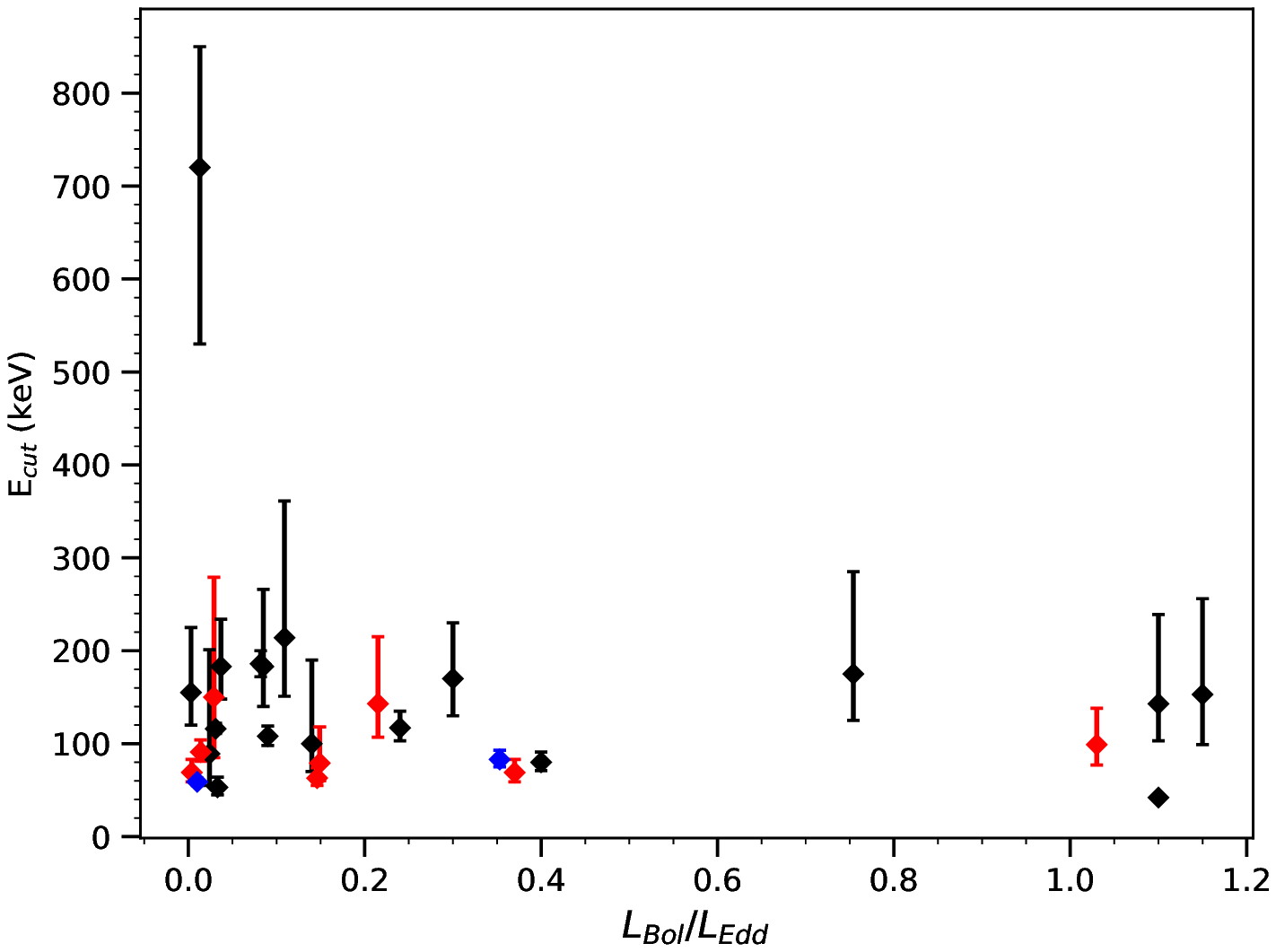}
    }
\caption{\label{Corr}Correlation between $E_{cut}$ and  $M_{BH}$ (top panel), $E_{cut}$ and  $\Gamma$ (middle panel) and $E_{cut}$ and  $\lambda_{Edd}$ (bottom panel).
The red points belong to the sources analysed in this work, the two blue points are from our earlier work on two sources 3C 120 and NGC 4151, while the black points
are for the sources collected from literature. The green 
lines in the middle panel are the unweighted linear least squares fit to sources with $\Gamma < 1.78$ and 1.78 $<$ $\Gamma$ $<$ 2.0 respectively. }
\end{figure}

\begin{figure}
\hspace*{-0.5cm} \includegraphics[scale=0.6]{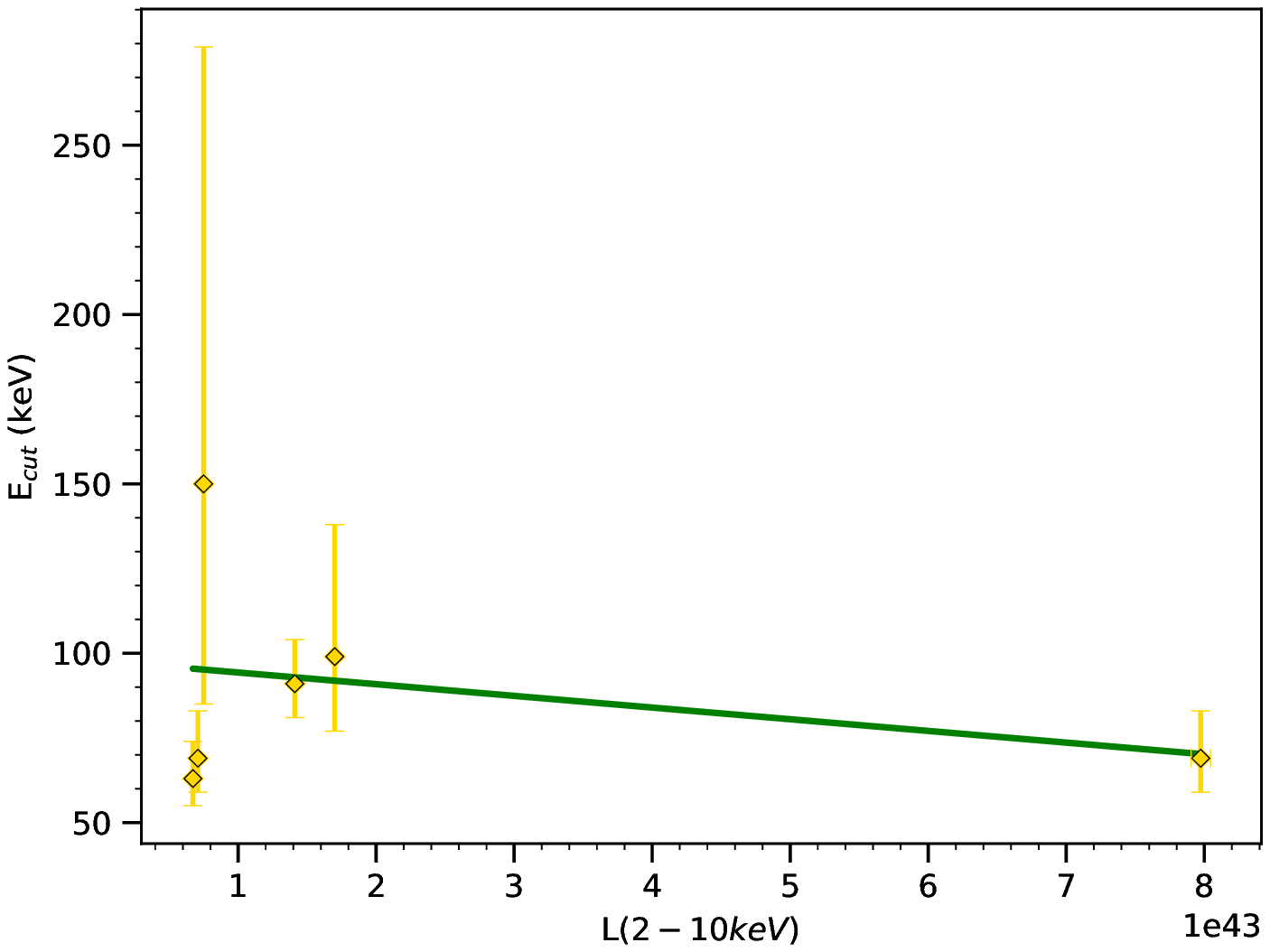}
\caption{\label{luminosity}$E_{cut}$ vs luminosity in the 2$-$10 keV band for sources with 1.78 $<$ $\Gamma$ $<$ 2.0 and showing a negative correlation between $E_{cut}$ and $\Gamma$ in the
middle panel of Fig. \ref{Corr}}
\end{figure}

\subsubsection{Model-3}
While the fits to the spectra using the model {\tt TBbs$\times$zTbabs$\times$(zgauss+pexrav)} is 
acceptable, we replaced the Gaussian component in Model-2 with
the relativistic line emission model {\tt RELLINE} \citep{2010MNRAS.409.1534D} 
and refit the spectra. The parameters obtained using {\tt RELLINE} model are similar 
to that obtained using {\tt TBbs$\times$zTbabs$\times$(zgauss+pexrav)} model. There is negligible 
improvement in the parameters obtained with Model-2 suggesting little/no 
blurring. Hence, in all further discussions we consider the parameters obtained 
by the model {\tt TBbs$\times$zTbabs$\times$(zgauss+pexrav).}
\subsubsection{Reflection parameter}
All the 10 sources studied in this work are Seyfert galaxies, however
based on \cite{2010A&A...518A..10V} they  have
varied classifications such as Sy1, Sy1.2, Sy1.5, Sy1.9 and Sy2. Clubbing
all sources with classifications up to Sy1.5 as Sy1 galaxies and sources beyond
Sy1.5 as Seyfert 2 galaxies, we found two Seyfert 2 galaxies and eight Seyfert 1 galaxies
The unweighted mean value of R for the Seyfert 2 galaxies  in our sample is 0.58 $\pm$ 0.51,
while that for the Seyfert 1 galaxy sample, we obtained an unweighted
mean value of 1.05 $\pm$ 0.66. Given the large error bars, both
Sy1 and Sy2 galaxies have similar  mean $R$ value, however, this large
error bar is attributable to the small number statistics. Given this
limitation, the mean value of R for Seyfert 2 galaxies points to have a lower value
compared to the mean R value of Seyfert 1 galaxies. The decrease of reflection in
Seyfert 2 relative to Seyfert 1 galaxies would be in agreement with the Unification scenario
\citep{1995PASP..107..803U}. Reprocessing in AGN is from the accretion disk and
for Seyfert 1 galaxies that are observed pole on, we are able to see more of the
reprocessed radiation, while in Seyfert 2 galaxies that are observed edge on,
the reprocessed component is expected to be less. From an analysis of Swift/BAT
spectra for a large sample of AGN, \cite{2017ApJS..233...17R} found obscured sources
to have less values of $R$ compared  to their counterparts that are seen pole on.Thus our results on $R$, though suffer from small number statistics are
in agreement with that found by \cite{2017ApJS..233...17R} from an analysis of  the
spectra taken from Swift/BAT for a  larger number of sources.
However, from an  analysis of the stacked Swift/BAT spectra, \cite{2013ApJ...770L..37V} found
that obscured sources have more  reflection component than their unobscured
counterparts. The origin of this difference between the values obtained
from spectral analysis of individual sources and analysis of the stacked
spectra of different categories of sources is not clear.

\subsubsection{Photon index}
The photon indices obtained by both the model fits ranged between 1.57 to 2.03.
Comparing the photon indices obtained from both the model fits, we noticed that
the $\Gamma$ obtained by model-1 (a simple power law fit) is flatter than the
$\Gamma$ obtained from model-2 for all the sources except 1, namely
NGC 2992. The steeper $\Gamma$ obtained from model-2 is also consistent with
the observations of the presence of high energy cut-off in most of the AGN. Unweighted mean values obtained from both model-1 and model-2 are
1.77 $\pm$ 0.12 and 1.86 $\pm$ 0.10 respectively. The plot of the $\Gamma$ obtained from model-1 against $\Gamma$ obtained from model-2 is shown in
Fig. \ref{index}. Also, shown in the same figure is a line of unity slope. It is very clear from the Figure, that the $\Gamma$ from model-2 is steeper than
the $\Gamma$ obtained from model-1.

\subsubsection{Cut-off energy}
Of the 10 sources analysed here, we obtained $E_{cut}$ for 8 sources, for one
source a lower limit is obtained while for one source, we could not constrain
$E_{cut}$. For sources for which we were able to obtain $E_{cut}$, the obtained
values range between 160 keV $<$ $E_{cut}$ $<$ 63 keV. For 4 sources in our sample, the obtained
$E_{cut}$ values were less than 80 keV and is within the energy range for which
{\it NuSTAR} is sensitive. For our sample of 8 sources, we found a mean $E_{cut}$ value
of 95 keV  with a standard deviation of 32 keV. This is lower than that
obtained by \cite{2014ApJ...782L..25M}, who on analysis of 41 AGN found a mean
$E_{cut}$ value of 128 keV and a standard deviation of 46 keV. This comparison needs to be
taken with caution as changes in the $E_{cut}$ values, that reflect coronal 
temperature variations are also noticed for
sources when observed at different times \citep{2017ApJ...836....2Z,2018ApJ...863...71Z}.

\subsection{Correlation of $E_{cut}$ with other parameters}
By modelling the observed X-ray spectra of 10 AGN using data from {\it NuSTAR}
using an empirical description of the observations as a power law with
an exponential cut-off, we were able to derive $\Gamma$ for 10 sources.
Out of the 10 sources, we could obtain $E_{cut}$ for 8 sources, and a lower limit for
one source. Using these new measurements along with data for other sources
culled from literature that has {\it NuSTAR} measurements, we could collect data
for a total of 30 sources. The $\Gamma$ values for this enlarged complete sample,  range from 1.6 to 2.4, while the
$E_{cut}$ take values lesser then 250 keV, except for one sources namely NGC 5506 having a
value of $E_{cut} = 720^{+130}_{-190}$.  This range of $E_{cut}$ from {\it NuSTAR} also lies in the range of
$E_{cut}$ values obtained from non-focussing instruments such as BeppoSAX and
INTEGRAL. However, the values of $E_{cut}$ from {\it NuSTAR} have low errors compared to
the values obtained from earlier missions operating in the energy range similar
to {\it NuSTAR}. This is likely due to the high sensitivity of {\it NuSTAR} compared to
earlier missions.  For these 30 sources with quality $E_{cut}$ measurements from
{\it NusTAR}, we tried to look for correlation
if any between $E_{cut}$ and various properties of the sources, such as 
$\Gamma$, BH mass and Eddington ratio. We obtained a complicated pattern between
$E_{cut}$ and $\Gamma$. This is shown in Fig. \ref{Corr}. For sources with $\Gamma$ less than 1.78, we found
a positive correlation (correlation coefficient = 0.6) between $E_{cut}$ and $\Gamma$, while if we consider
sources with 1.78 < $\Gamma$ < 2.0, we found a negative correlation (correlation coefficient = 0.6)  between
$E_{cut}$ and $\Gamma$. Beyond $\Gamma$ > 2.0, no trend of $E_{cut}$ with $\Gamma$ is noticed,
however, this apparent no-correlation is based on three sources. Thus this
analysis gives indications of  the existence of complicated correlation
between $E_{cut}$ and $\Gamma$.  Though the reasons for this complicated
behaviour is not clear presently, the existence of it too  needs to be 
confirmed from more precise
measurements of $E_{cut}$ on a larger number of sources. For the sources
lying in the negative correlation line in the $E_{cut}$ versus $\Gamma$ diagram, we
plot in Fig. \ref{luminosity} the $E_{cut}$ of those sources against their luminosity in the 
2$-$10 keV band. We noticed a weak
negative correlation with a correlation coefficient of 0.3  wherein sources with
low $E_{cut}$ have larger luminosity. This behaviour can be explained due to electrons in 
the corona being more effectively cooled 
via Comptonization in luminous sources, thereby leading to low $E_{cut}$ as well
as steeper $\Gamma$ \citep{2018ApJ...863...71Z}.  From BeppoSAX measurement of nine sources, using
data in the range of 0.1 $-$ 200 keV, \cite{2002A&A...389..802P} found for the
first time a  strong positive correlation between $E_{cut}$ and $\Gamma$. In their
sample of nine sources, two have lower limits and some from the remaining seven have large error bars.  From
simulated Swift/BAT data \cite{2017ApJS..233...17R} found a negative correlation between
$E_{cut}$ and $\Gamma$ while \cite{2018A&A...614A..37T} using a sample of 19 sources,
found no correlation between $E_{cut}$ and $\Gamma$. We also looked for
correlation between $E_{cut}$ and Eddington ratio ($\lambda_{Edd}$ = $L_{Bol}$/$L_{Edd}$).
To estimate $L_{Bol}$ for our sources we calculated the  intrinsic (absorption corrected and
$k$-corrected) continuum luminosity in 2$-$ 10 keV using the following relation

\begin{equation}
L_{int} = 4 \pi d_L^2 \frac{F_{int}}{(1 + z)^{2 - \Gamma}}
\end{equation}

where $F_{int}$ is the absorption corrected 2$-$10 keV flux and $d_L$ is the
luminosity distance. From $L_{int}$, $L_{Bol}$ was calculated as
$L_{Bol} = 20 \times L_{int}$ \citep{2007MNRAS.381.1235V}. We did not 
find any correlation between $E_{cut}$ and Eddington ratio.
The correlation between $E_{cut}$ and BH mass is shown in the top panel of Fig. \ref{Corr}. Also, shown in the same figure
are unweighted linear least squares fit (magenta line) and weighted linear least squares fit (yellow line). There is an 
indication of a weak positive correlation.

Recently, \cite{2018A&A...614A..37T} found an anti-correlation between the 
coronal temperature and optical depth ($\tau$) from an analysis of a sample of 
Seyfert galaxies. We in this work have first time  measurement 
of $E_{cut}$ for eight Seyfert galaxies. We tried to investigate the location 
of our eight new sources in the $KT_e$ $-$ $\tau$ plane and see if they
lie on the trend  found by \cite{2018A&A...614A..37T}. To calculate 
$\tau$ we used the approximation given by \cite{1979PAZh....5..279P} as

\begin{equation}
\Gamma = 1 + \frac{[2/(\theta + 3) - log (\tau)]}  {log (12 \theta^2 + 25 \theta)}
\end{equation}

\begin{figure}
\hspace*{-0.5cm} \includegraphics[scale=0.6]{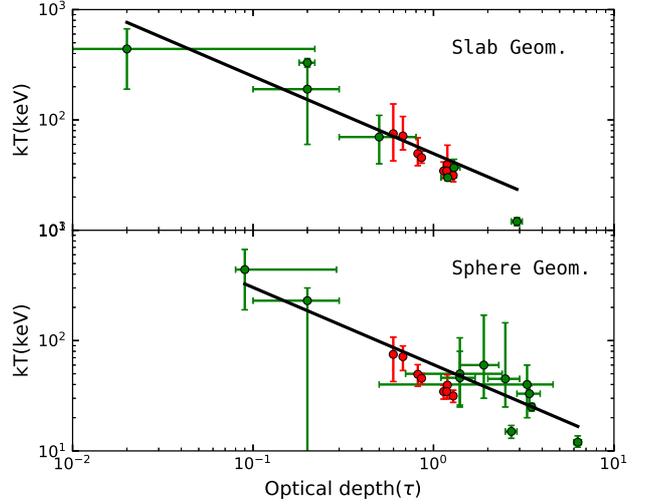}
\caption{\label{kte} Coronal temperature versus optical depth for Seyfert 
galaxies in the case of slab geometry (top panel) and spherical geometry 
(bottom panel). The green filled circles are the measurements from 
\citet{2018A&A...614A..37T} while the red filled circles are the new 
measurements from this work.  The black solid lines are the relation 
from \citet{2018A&A...614A..37T} separately for the disk and 
spherical shape of the corona.}

\end{figure}

Similarly for $KT_e$, we used $KT_e = E_{cut}/2$ \citep{2001ApJ...556..716P}. 
We show in Fig. \ref{kte} the location of our sources in the $KT_e$ versus 
$\tau$ plane both for slab and spherical geometry of the corona. Also, shown
in the same figure are the sources with $KT_e$ measurements from
\cite{2018A&A...614A..37T} as well as the relation found by
\cite{2018A&A...614A..37T} separately for the slab and spherical geometry. Our
sources nicely lie in the trend found by \cite{2018A&A...614A..37T}.

\begin{figure}
\hspace*{-0.5cm} \includegraphics[scale=0.6]{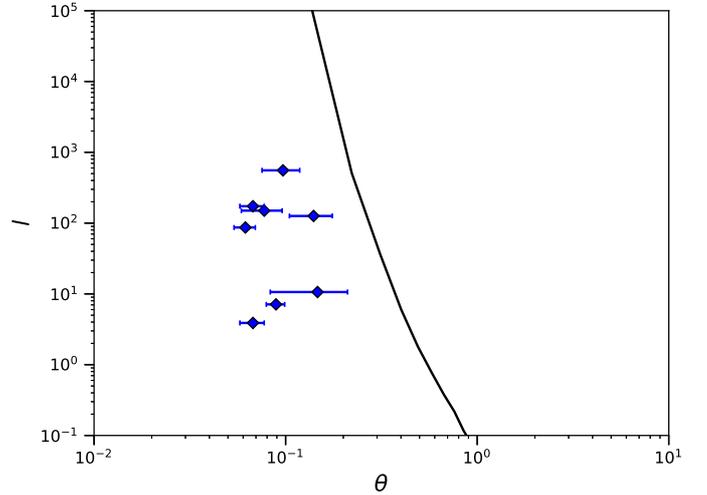}
\caption{\label{thetalplane} Location of our sources in the $\theta$ $-$ $l$ plane.
The black solid line corresponds to the pair line for the slab
coronal geometry.}.

\end{figure}

\subsection{Location of sources in the $\theta$ $-$ $l$ plane}
We have E$_{cut}$ measurements for eight sources. To plot the location of our 
sources in the $\theta$ - $l$ plane we converted out E$_{cut}$ measurements to 
$\theta$ using Equation (2) given in Section 1, where we 
used $K_B T_e$ = $E_{cut}/2$ \citep{2001ApJ...556..716P}. Similarly for calculating $l$ we used 
Equation (1). Here, for the coronal radius we assumed a value of 10$R_G$ 
\citep{2015MNRAS.451.4375F}, as we do not have any measurement of the coronal 
size for our sources. For the luminosity
of the sources, we used the absorption corrected  
0.1$-$200 keV flux  obtained from our spectral fits and converted to luminosity
using the luminosity distance. Black hole masses for the sources were taken from literature. We show in 
Fig. \ref{thetalplane} the location of our sources in the $\theta$ $-$ $l$ plane. Also shown in the same diagram is the pair line
for a slab geometry \citep{1995ApJ...449L..13S,2015MNRAS.451.4375F}. All the sources for which $E_{cut}$ has been derived in this work lie within the 
theoretical pair line, similar to that found by \cite{2015MNRAS.451.4375F}  
and \cite{2018ApJ...866..124K}. 

\section{Summary and conclusion}
We have carried out X-ray spectral and timing analysis of a sample of 10 sources, using data from {\it NuSTAR}. The aims of this work are
two fold (a) to provide new measurements of $E_{cut}$  in AGN and (b) look for correlations between $E_{cut}$ values obtained only from 
{\it NuSTAR data}  and other physical parameters of the sources. The results of this work are summarized below

\begin{enumerate}
\item All the sources showed flux variations in, soft, hard and the total bands. For 6 out of 10 sources, the amplitude of 
flux variations in the soft band is larger than that of the hard band. Three sources, namely Mrk 348, NGC 2992 and NGC 7172 show similar variation
in both hard and soft bands, while in one source, Mrk 509, the variation in the hard band is larger than that of the soft band.
Though differences in flux variability between soft and hard bands were noticed in individual sources, for the complete sample analysed here
statistically there is no difference between the flux variations in the hard and soft bands.

\item The HR was used as a proxy to characterise the spectra of the sources. No correlation of the HR with flux variations in
the total band was found for the sources, except one,  indicating that the spectra were non-variable during {\it NuSTAR} observations.
For NGC 7314, we found a softer when brighter trend  with a correlation coefficient of 0.53.

\item In seven out of 10 sources, FeK$\alpha$ line was found, while for three sources, namely Mrk 348, NGC 2992 and NGC 7172, FeK$\alpha$ line
could not be seen in their spectra.

\item Among the ten sources whose spectra were analysed, $E_{cut}$ values were obtained for eight sources. For one sources,
ESO 362-G18, a lower limit to the $E_{cut}$  value was estimated, while for NGC 7314, our spectral fits did not yield any $E_{cut}$ value.

\item Using the new $E_{cut}$ values obtained in this work along with those collected from literature, we could gather $E_{cut}$ measurements
for 30 sources. In this enlarged sample of 30 sources, we found no correlation between $E_{cut}$ and $M_{BH}$ and $E_{cut}$ and 
$\lambda_{Edd}$. However, we noticed a complicated correlation between $E_{cut}$ and $\Gamma$. For values of $\Gamma$ less than 1.78, 
$E_{cut}$ is positively correlated with $\Gamma$, while for $\Gamma$ values between 1.78 and 2.0, $E_{cut}$ is negatively correlated
with $\Gamma$.
\end{enumerate}

Though there has been an increase in the number of AGN with $E_{cut}$ measurements
from {\it NuSTAR}, it is still insufficient. Therefore, to study various
correlations and to put any constraints on the theory based on observations
the number of $E_{cut}$ measurements need to be increased. This also requires
physical model fits to the observed data to infer many other parameters of the system, rather
than phenomenological model fits, requiring high quality data from {\it NuSTAR}.

% 
% %%%%%%%%%%%%%%%%%%%%%%%%%%%%%%%%%%%%%%%%%%%%%%%%%%%%%%%%%%%%%%%%%%%%%%%%%%%%%%%%%%%%%%%%%%%%%%%%%%%%%%%%%%%%%%%%%%%%
%--------------------------------------------------------------------------------------------------------------

\section*{Acknowledgements}
This research made use of data from the {\it NuSTAR} mission, a project 
led by the 
California Institute of Technology, managed by the Jet Propulsion Laboratory, 
and funded by NASA, XMM-Newton, an ESA science mission with instruments and 
contributions directly funded by ESA Member States and NASA. This research has 
made use of the {\it NuSTAR} Data Analysis Software (NuSTARDAS) jointly developed by 
the ASI Science Data Center (ASDC, Italy) and the California Institute of 
Technology (USA).

\bibliographystyle{mnras}
\bibliography{master.bib}

\begin{thebibliography}{}
\makeatletter
\relax
\def\mn@urlcharsother{\let\do\@makeother \do\$\do\&\do\#\do\^\do\_\do\%\do\~}
\def\mn@doi{\begingroup\mn@urlcharsother \@ifnextchar [ {\mn@doi@}
  {\mn@doi@[]}}
\def\mn@doi@[#1]#2{\def\@tempa{#1}\ifx\@tempa\@empty \href
  {http://dx.doi.org/#2} {doi:#2}\else \href {http://dx.doi.org/#2} {#1}\fi
  \endgroup}
\def\mn@eprint#1#2{\mn@eprint@#1:#2::\@nil}
\def\mn@eprint@arXiv#1{\href {http://arxiv.org/abs/#1} {{\tt arXiv:#1}}}
\def\mn@eprint@dblp#1{\href {http://dblp.uni-trier.de/rec/bibtex/#1.xml}
  {dblp:#1}}
\def\mn@eprint@#1:#2:#3:#4\@nil{\def\@tempa {#1}\def\@tempb {#2}\def\@tempc
  {#3}\ifx \@tempc \@empty \let \@tempc \@tempb \let \@tempb \@tempa \fi \ifx
  \@tempb \@empty \def\@tempb {arXiv}\fi \@ifundefined
  {mn@eprint@\@tempb}{\@tempb:\@tempc}{\expandafter \expandafter \csname
  mn@eprint@\@tempb\endcsname \expandafter{\@tempc}}}

\bibitem[\protect\citeauthoryear{{Anders} \& {Grevesse}}{{Anders} \&
  {Grevesse}}{1989}]{1989GeCoA..53..197A}
{Anders} E.,  {Grevesse} N.,  1989, \mn@doi [\gca]
  {10.1016/0016-7037(89)90286-X}, \href
  {http://adsabs.harvard.edu/abs/1989GeCoA..53..197A} {53, 197}

\bibitem[\protect\citeauthoryear{{Arnaud}}{{Arnaud}}{1996}]{1996ASPC..101...17A}
{Arnaud} K.~A.,  1996, in {Jacoby} G.~H.,  {Barnes} J.,  eds,  Astronomical
  Society of the Pacific Conference Series Vol. 101, Astronomical Data Analysis
  Software and Systems V. p.~17

\bibitem[\protect\citeauthoryear{{Ballantyne} et~al.,}{{Ballantyne}
  et~al.}{2014}]{2014ApJ...794...62B}
{Ballantyne} D.~R.,  et~al., 2014, \mn@doi [\apj] {10.1088/0004-637X/794/1/62},
  \href {http://adsabs.harvard.edu/abs/2014ApJ...794...62B} {794, 62}

\bibitem[\protect\citeauthoryear{{Balokovi{\'c}} et~al.,}{{Balokovi{\'c}}
  et~al.}{2015}]{2015ApJ...800...62B}
{Balokovi{\'c}} M.,  et~al., 2015, \mn@doi [\apj] {10.1088/0004-637X/800/1/62},
  \href {http://adsabs.harvard.edu/abs/2015ApJ...800...62B} {800, 62}

\bibitem[\protect\citeauthoryear{{Balucinska-Church} \&
  {McCammon}}{{Balucinska-Church} \& {McCammon}}{1992}]{1992ApJ...400..699B}
{Balucinska-Church} M.,  {McCammon} D.,  1992, \mn@doi [\apj] {10.1086/172032},
  \href {http://adsabs.harvard.edu/abs/1992ApJ...400..699B} {400, 699}

\bibitem[\protect\citeauthoryear{{Bhayani} \& {Nandra}}{{Bhayani} \&
  {Nandra}}{2011}]{2011MNRAS.416..629B}
{Bhayani} S.,  {Nandra} K.,  2011, \mn@doi [\mnras]
  {10.1111/j.1365-2966.2011.19073.x}, \href
  {http://adsabs.harvard.edu/abs/2011MNRAS.416..629B} {416, 629}

\bibitem[\protect\citeauthoryear{{Brenneman} et~al.,}{{Brenneman}
  et~al.}{2014}]{2014ApJ...781...83B}
{Brenneman} L.~W.,  et~al., 2014, \mn@doi [\apj] {10.1088/0004-637X/781/2/83},
  \href {http://adsabs.harvard.edu/abs/2014ApJ...781...83B} {781, 83}

\bibitem[\protect\citeauthoryear{{Buisson}, {Fabian}  \& {Lohfink}}{{Buisson}
  et~al.}{2018}]{2018MNRAS.481.4419B}
{Buisson} D.~J.~K.,  {Fabian} A.~C.,   {Lohfink} A.~M.,  2018, \mn@doi [\mnras]
  {10.1093/mnras/sty2609}, \href
  {http://adsabs.harvard.edu/abs/2018MNRAS.481.4419B} {481, 4419}

\bibitem[\protect\citeauthoryear{{Chartas}, {Kochanek}, {Dai}, {Poindexter}  \&
  {Garmire}}{{Chartas} et~al.}{2009}]{2009ApJ...693..174C}
{Chartas} G.,  {Kochanek} C.~S.,  {Dai} X.,  {Poindexter} S.,   {Garmire} G.,
  2009, \mn@doi [\apj] {10.1088/0004-637X/693/1/174}, \href
  {http://adsabs.harvard.edu/abs/2009ApJ...693..174C} {693, 174}

\bibitem[\protect\citeauthoryear{{Dadina}}{{Dadina}}{2007}]{2007A&A...461.1209D}
{Dadina} M.,  2007, \mn@doi [\aap] {10.1051/0004-6361:20065734}, \href
  {http://adsabs.harvard.edu/abs/2007A%26A...461.1209D} {461, 1209}

\bibitem[\protect\citeauthoryear{{Dauser}, {Wilms}, {Reynolds}  \&
  {Brenneman}}{{Dauser} et~al.}{2010}]{2010MNRAS.409.1534D}
{Dauser} T.,  {Wilms} J.,  {Reynolds} C.~S.,   {Brenneman} L.~W.,  2010,
  \mn@doi [\mnras] {10.1111/j.1365-2966.2010.17393.x}, \href
  {http://adsabs.harvard.edu/abs/2010MNRAS.409.1534D} {409, 1534}

\bibitem[\protect\citeauthoryear{{Dickey} \& {Lockman}}{{Dickey} \&
  {Lockman}}{1990}]{1990ARA&A..28..215D}
{Dickey} J.~M.,  {Lockman} F.~J.,  1990, \mn@doi [\araa]
  {10.1146/annurev.aa.28.090190.001243}, \href
  {http://adsabs.harvard.edu/abs/1990ARA%26A..28..215D} {28, 215}

\bibitem[\protect\citeauthoryear{{Dove}, {Wilms}, {Maisack}  \&
  {Begelman}}{{Dove} et~al.}{1997}]{1997ApJ...487..759D}
{Dove} J.~B.,  {Wilms} J.,  {Maisack} M.,   {Begelman} M.~C.,  1997, \mn@doi
  [\apj] {10.1086/304647}, \href
  {http://adsabs.harvard.edu/abs/1997ApJ...487..759D} {487, 759}

\bibitem[\protect\citeauthoryear{{Edelson}, {Turner}, {Pounds}, {Vaughan},
  {Markowitz}, {Marshall}, {Dobbie}  \& {Warwick}}{{Edelson}
  et~al.}{2002}]{2002ApJ...568..610E}
{Edelson} R.,  {Turner} T.~J.,  {Pounds} K.,  {Vaughan} S.,  {Markowitz} A.,
  {Marshall} H.,  {Dobbie} P.,   {Warwick} R.,  2002, \mn@doi [\apj]
  {10.1086/323779}, \href {http://adsabs.harvard.edu/abs/2002ApJ...568..610E}
  {568, 610}

\bibitem[\protect\citeauthoryear{{Fabian}}{{Fabian}}{1999}]{1999PNAS...96.4749F}
{Fabian} A.~C.,  1999, \mn@doi [Proceedings of the National Academy of Science]
  {10.1073/pnas.96.9.4749}, \href
  {http://adsabs.harvard.edu/abs/1999PNAS...96.4749F} {96, 4749}

\bibitem[\protect\citeauthoryear{{Fabian}, {Ballantyne}, {Merloni}, {Vaughan},
  {Iwasawa}  \& {Boller}}{{Fabian} et~al.}{2002}]{2002MNRAS.331L..35F}
{Fabian} A.~C.,  {Ballantyne} D.~R.,  {Merloni} A.,  {Vaughan} S.,  {Iwasawa}
  K.,   {Boller} T.,  2002, \mn@doi [\mnras]
  {10.1046/j.1365-8711.2002.05419.x}, \href
  {http://adsabs.harvard.edu/abs/2002MNRAS.331L..35F} {331, L35}

\bibitem[\protect\citeauthoryear{{Fabian} et~al.,}{{Fabian}
  et~al.}{2009}]{2009Natur.459..540F}
{Fabian} A.~C.,  et~al., 2009, \mn@doi [\nat] {10.1038/nature08007}, \href
  {http://adsabs.harvard.edu/abs/2009Natur.459..540F} {459, 540}

\bibitem[\protect\citeauthoryear{{Fabian}, {Lohfink}, {Kara}, {Parker},
  {Vasudevan}  \& {Reynolds}}{{Fabian} et~al.}{2015}]{2015MNRAS.451.4375F}
{Fabian} A.~C.,  {Lohfink} A.,  {Kara} E.,  {Parker} M.~L.,  {Vasudevan} R.,
  {Reynolds} C.~S.,  2015, \mn@doi [\mnras] {10.1093/mnras/stv1218}, \href
  {http://adsabs.harvard.edu/abs/2015MNRAS.451.4375F} {451, 4375}

\bibitem[\protect\citeauthoryear{{Guilbert}, {Fabian}  \& {Rees}}{{Guilbert}
  et~al.}{1983}]{1983MNRAS.205..593G}
{Guilbert} P.~W.,  {Fabian} A.~C.,   {Rees} M.~J.,  1983, \mn@doi [\mnras]
  {10.1093/mnras/205.3.593}, \href
  {http://adsabs.harvard.edu/abs/1983MNRAS.205..593G} {205, 593}

\bibitem[\protect\citeauthoryear{{Haardt} \& {Maraschi}}{{Haardt} \&
  {Maraschi}}{1991}]{1991ApJ...380L..51H}
{Haardt} F.,  {Maraschi} L.,  1991, \mn@doi [\apjl] {10.1086/186171}, \href
  {http://adsabs.harvard.edu/abs/1991ApJ...380L..51H} {380, L51}

\bibitem[\protect\citeauthoryear{{Haardt}, {Maraschi}  \&
  {Ghisellini}}{{Haardt} et~al.}{1994}]{1994ApJ...432L..95H}
{Haardt} F.,  {Maraschi} L.,   {Ghisellini} G.,  1994, \mn@doi [\apjl]
  {10.1086/187520}, \href {http://adsabs.harvard.edu/abs/1994ApJ...432L..95H}
  {432, L95}

\bibitem[\protect\citeauthoryear{{Haardt}, {Maraschi}  \&
  {Ghisellini}}{{Haardt} et~al.}{1997}]{1997ApJ...476..620H}
{Haardt} F.,  {Maraschi} L.,   {Ghisellini} G.,  1997, \mn@doi [\apj]
  {10.1086/303656}, \href {http://adsabs.harvard.edu/abs/1997ApJ...476..620H}
  {476, 620}

\bibitem[\protect\citeauthoryear{{Harrison} et~al.,}{{Harrison}
  et~al.}{2013}]{2013ApJ...770..103H}
{Harrison} F.~A.,  et~al., 2013, \mn@doi [\apj] {10.1088/0004-637X/770/2/103},
  \href {http://adsabs.harvard.edu/abs/2013ApJ...770..103H} {770, 103}

\bibitem[\protect\citeauthoryear{{Johnson}, {McNaron-Brown}, {Kurfess},
  {Zdziarski}, {Magdziarz}  \& {Gehrels}}{{Johnson}
  et~al.}{1997}]{1997ApJ...482..173J}
{Johnson} W.~N.,  {McNaron-Brown} K.,  {Kurfess} J.~D.,  {Zdziarski} A.~A.,
  {Magdziarz} P.,   {Gehrels} N.,  1997, \mn@doi [\apj] {10.1086/304148}, \href
  {http://adsabs.harvard.edu/abs/1997ApJ...482..173J} {482, 173}

\bibitem[\protect\citeauthoryear{{Kamraj}, {Harrison}, {Balokovi{\'c}},
  {Lohfink}  \& {Brightman}}{{Kamraj} et~al.}{2018}]{2018ApJ...866..124K}
{Kamraj} N.,  {Harrison} F.~A.,  {Balokovi{\'c}} M.,  {Lohfink} A.,
  {Brightman} M.,  2018, \mn@doi [\apj] {10.3847/1538-4357/aadd0d}, \href
  {http://adsabs.harvard.edu/abs/2018ApJ...866..124K} {866, 124}

\bibitem[\protect\citeauthoryear{{Kara}, {Fabian}, {Cackett}, {Uttley},
  {Wilkins}  \& {Zoghbi}}{{Kara} et~al.}{2013}]{2013MNRAS.434.1129K}
{Kara} E.,  {Fabian} A.~C.,  {Cackett} E.~M.,  {Uttley} P.,  {Wilkins} D.~R.,
  {Zoghbi} A.,  2013, \mn@doi [\mnras] {10.1093/mnras/stt1055}, \href
  {http://adsabs.harvard.edu/abs/2013MNRAS.434.1129K} {434, 1129}

\bibitem[\protect\citeauthoryear{{Kara}, {Garcia}, {Lohfink}, {Fabian},
  {Reynolds}, {Tombesi}  \& {Wilkins}}{{Kara}
  et~al.}{2017}]{2017arXiv170309815K}
{Kara} E.,  {Garcia} J.~A.,  {Lohfink} A.,  {Fabian} A.~C.,  {Reynolds} C.~S.,
  {Tombesi} F.,   {Wilkins} D.~R.,  2017, preprint, \href
  {http://adsabs.harvard.edu/abs/2017arXiv170309815K} {} (\mn@eprint {arXiv}
  {1703.09815})

\bibitem[\protect\citeauthoryear{Lanzuisi et~al.}{Lanzuisi
  et~al.}{2016}]{2016A&A...590A..77L}
Lanzuisi G.,  et~al., 2016, \mn@doi [Astron. Astrophys.]
  {10.1051/0004-6361/201628325}, 590, A77

\bibitem[\protect\citeauthoryear{{Lohfink} et~al.,}{{Lohfink}
  et~al.}{2015}]{2015ApJ...814...24L}
{Lohfink} A.~M.,  et~al., 2015, \mn@doi [\apj] {10.1088/0004-637X/814/1/24},
  \href {http://adsabs.harvard.edu/abs/2015ApJ...814...24L} {814, 24}

\bibitem[\protect\citeauthoryear{{Lohfink} et~al.,}{{Lohfink}
  et~al.}{2017}]{2017arXiv170403673L}
{Lohfink} A.,  et~al., 2017, preprint, \href
  {http://adsabs.harvard.edu/abs/2017arXiv170403673L} {} (\mn@eprint {arXiv}
  {1704.03673})

\bibitem[\protect\citeauthoryear{{Lubi{\'n}ski}, {Zdziarski}, {Walter},
  {Paltani}, {Beckmann}, {Soldi}, {Ferrigno}  \& {Courvoisier}}{{Lubi{\'n}ski}
  et~al.}{2010}]{2010MNRAS.408.1851L}
{Lubi{\'n}ski} P.,  {Zdziarski} A.~A.,  {Walter} R.,  {Paltani} S.,  {Beckmann}
  V.,  {Soldi} S.,  {Ferrigno} C.,   {Courvoisier} T.~J.-L.,  2010, \mn@doi
  [\mnras] {10.1111/j.1365-2966.2010.17251.x}, \href
  {http://adsabs.harvard.edu/abs/2010MNRAS.408.1851L} {408, 1851}

\bibitem[\protect\citeauthoryear{{Lubi{\'n}ski} et~al.,}{{Lubi{\'n}ski}
  et~al.}{2016}]{2016MNRAS.458.2454L}
{Lubi{\'n}ski} P.,  et~al., 2016, \mn@doi [\mnras] {10.1093/mnras/stw454},
  \href {http://adsabs.harvard.edu/abs/2016MNRAS.458.2454L} {458, 2454}

\bibitem[\protect\citeauthoryear{{Magdziarz} \& {Zdziarski}}{{Magdziarz} \&
  {Zdziarski}}{1995}]{1995MNRAS.273..837M}
{Magdziarz} P.,  {Zdziarski} A.~A.,  1995, \mn@doi [\mnras]
  {10.1093/mnras/273.3.837}, \href
  {http://adsabs.harvard.edu/abs/1995MNRAS.273..837M} {273, 837}

\bibitem[\protect\citeauthoryear{{Malizia}, {Molina}, {Bassani}, {Stephen},
  {Bazzano}, {Ubertini}  \& {Bird}}{{Malizia}
  et~al.}{2014}]{2014ApJ...782L..25M}
{Malizia} A.,  {Molina} M.,  {Bassani} L.,  {Stephen} J.~B.,  {Bazzano} A.,
  {Ubertini} P.,   {Bird} A.~J.,  2014, \mn@doi [\apjl]
  {10.1088/2041-8205/782/2/L25}, \href
  {http://adsabs.harvard.edu/abs/2014ApJ...782L..25M} {782, L25}

\bibitem[\protect\citeauthoryear{{Matt} et~al.,}{{Matt}
  et~al.}{2015}]{2015MNRAS.447.3029M}
{Matt} G.,  et~al., 2015, \mn@doi [\mnras] {10.1093/mnras/stu2653}, \href
  {http://adsabs.harvard.edu/abs/2015MNRAS.447.3029M} {447, 3029}

\bibitem[\protect\citeauthoryear{{McHardy}, {Gunn}, {Uttley}  \&
  {Goad}}{{McHardy} et~al.}{2005}]{2005MNRAS.359.1469M}
{McHardy} I.~M.,  {Gunn} K.~F.,  {Uttley} P.,   {Goad} M.~R.,  2005, \mn@doi
  [\mnras] {10.1111/j.1365-2966.2005.08992.x}, \href
  {http://adsabs.harvard.edu/abs/2005MNRAS.359.1469M} {359, 1469}

\bibitem[\protect\citeauthoryear{{Middei} et~al.,}{{Middei}
  et~al.}{2018}]{2018A&A...615A.163M}
{Middei} R.,  et~al., 2018, \mn@doi [\aap] {10.1051/0004-6361/201832726}, \href
  {http://adsabs.harvard.edu/abs/2018A%26A...615A.163M} {615, A163}

\bibitem[\protect\citeauthoryear{{Mushotzky}, {Done}  \& {Pounds}}{{Mushotzky}
  et~al.}{1993}]{1993ARA&A..31..717M}
{Mushotzky} R.~F.,  {Done} C.,   {Pounds} K.~A.,  1993, \mn@doi [\araa]
  {10.1146/annurev.astro.31.1.717}, \href
  {http://adsabs.harvard.edu/abs/1993ARA%26A..31..717M} {31, 717}

\bibitem[\protect\citeauthoryear{{Nicastro} et~al.,}{{Nicastro}
  et~al.}{2000}]{2000ApJ...536..718N}
{Nicastro} F.,  et~al., 2000, \mn@doi [\apj] {10.1086/308950}, \href
  {http://adsabs.harvard.edu/abs/2000ApJ...536..718N} {536, 718}

\bibitem[\protect\citeauthoryear{{Perola}, {Matt}, {Cappi}, {Fiore},
  {Guainazzi}, {Maraschi}, {Petrucci}  \& {Piro}}{{Perola}
  et~al.}{2002}]{2002A&A...389..802P}
{Perola} G.~C.,  {Matt} G.,  {Cappi} M.,  {Fiore} F.,  {Guainazzi} M.,
  {Maraschi} L.,  {Petrucci} P.~O.,   {Piro} L.,  2002, \mn@doi [\aap]
  {10.1051/0004-6361:20020658}, \href
  {http://adsabs.harvard.edu/abs/2002A%26A...389..802P} {389, 802}

\bibitem[\protect\citeauthoryear{{Petrucci} et~al.,}{{Petrucci}
  et~al.}{2001}]{2001ApJ...556..716P}
{Petrucci} P.~O.,  et~al., 2001, \mn@doi [\apj] {10.1086/321629}, \href
  {http://adsabs.harvard.edu/abs/2001ApJ...556..716P} {556, 716}

\bibitem[\protect\citeauthoryear{{Petrucci} et~al.,}{{Petrucci}
  et~al.}{2013}]{2013A&A...549A..73P}
{Petrucci} P.-O.,  et~al., 2013, \mn@doi [\aap] {10.1051/0004-6361/201219956},
  \href {http://adsabs.harvard.edu/abs/2013A%26A...549A..73P} {549, A73}

\bibitem[\protect\citeauthoryear{Porquet et~al.}{Porquet
  et~al.}{2018}]{2018A&A...609A..42P}
Porquet D.,  et~al., 2018, \mn@doi [Astron. Astrophys.]
  {10.1051/0004-6361/201731290}, 609, A42

\bibitem[\protect\citeauthoryear{{Poutanen}, {Svensson}  \& {Stern}}{{Poutanen}
  et~al.}{1997}]{1997ESASP.382..401P}
{Poutanen} J.,  {Svensson} R.,   {Stern} B.,  1997, in {Winkler} C.,
  {Courvoisier} T.~J.-L.,   {Durouchoux} P.,  eds,  ESA Special Publication
  Vol. 382, The Transparent Universe. p.~401 (\mn@eprint {} {astro-ph/9701168})

\bibitem[\protect\citeauthoryear{{Pozdniakov}, {Sobol}  \&
  {Siuniaev}}{{Pozdniakov} et~al.}{1979}]{1979PAZh....5..279P}
{Pozdniakov} L.~A.,  {Sobol} I.~M.,   {Siuniaev} R.~A.,  1979, Pisma v
  Astronomicheskii Zhurnal, \href
  {http://adsabs.harvard.edu/abs/1979PAZh....5..279P} {5, 279}

\bibitem[\protect\citeauthoryear{{Rani} \& {Stalin}}{{Rani} \&
  {Stalin}}{2018a}]{2018JApA...39...15R}
{Rani} P.,  {Stalin} C.~S.,  2018a, \mn@doi [Journal of Astrophysics and
  Astronomy] {10.1007/s12036-017-9502-5}, \href
  {http://adsabs.harvard.edu/abs/2018JApA...39...15R} {39, 15}

\bibitem[\protect\citeauthoryear{{Rani} \& {Stalin}}{{Rani} \&
  {Stalin}}{2018b}]{2018ApJ...856..120R}
{Rani} P.,  {Stalin} C.~S.,  2018b, \mn@doi [\apj] {10.3847/1538-4357/aab356},
  \href {http://adsabs.harvard.edu/abs/2018ApJ...856..120R} {856, 120}

\bibitem[\protect\citeauthoryear{{Rees}}{{Rees}}{1984}]{1984ARA&A..22..471R}
{Rees} M.~J.,  1984, \mn@doi [\araa] {10.1146/annurev.aa.22.090184.002351},
  \href {http://adsabs.harvard.edu/abs/1984ARA%26A..22..471R} {22, 471}

\bibitem[\protect\citeauthoryear{{Ricci}, {Walter}, {Courvoisier}  \&
  {Paltani}}{{Ricci} et~al.}{2011}]{2011A&A...532A.102R}
{Ricci} C.,  {Walter} R.,  {Courvoisier} T.~J.-L.,   {Paltani} S.,  2011,
  \mn@doi [\aap] {10.1051/0004-6361/201016409}, \href
  {http://adsabs.harvard.edu/abs/2011A%26A...532A.102R} {532, A102}

\bibitem[\protect\citeauthoryear{{Ricci} et~al.,}{{Ricci}
  et~al.}{2017}]{2017ApJS..233...17R}
{Ricci} C.,  et~al., 2017, \mn@doi [\apjs] {10.3847/1538-4365/aa96ad}, \href
  {http://adsabs.harvard.edu/abs/2017ApJS..233...17R} {233, 17}

\bibitem[\protect\citeauthoryear{{Ricci} et~al.,}{{Ricci}
  et~al.}{2018}]{2018MNRAS.480.1819R}
{Ricci} C.,  et~al., 2018, \mn@doi [\mnras] {10.1093/mnras/sty1879}, \href
  {http://adsabs.harvard.edu/abs/2018MNRAS.480.1819R} {480, 1819}

\bibitem[\protect\citeauthoryear{{Risaliti}, {Elvis}, {Fabbiano}, {Baldi}  \&
  {Zezas}}{{Risaliti} et~al.}{2005}]{2005ApJ...623L..93R}
{Risaliti} G.,  {Elvis} M.,  {Fabbiano} G.,  {Baldi} A.,   {Zezas} A.,  2005,
  \mn@doi [\apjl] {10.1086/430252}, \href
  {http://adsabs.harvard.edu/abs/2005ApJ...623L..93R} {623, L93}

\bibitem[\protect\citeauthoryear{{Risaliti}, {Nardini}, {Salvati}, {Elvis},
  {Fabbiano}, {Maiolino}, {Pietrini}  \& {Torricelli-Ciamponi}}{{Risaliti}
  et~al.}{2011}]{2011MNRAS.410.1027R}
{Risaliti} G.,  {Nardini} E.,  {Salvati} M.,  {Elvis} M.,  {Fabbiano} G.,
  {Maiolino} R.,  {Pietrini} P.,   {Torricelli-Ciamponi} G.,  2011, \mn@doi
  [\mnras] {10.1111/j.1365-2966.2010.17503.x}, \href
  {http://adsabs.harvard.edu/abs/2011MNRAS.410.1027R} {410, 1027}

\bibitem[\protect\citeauthoryear{{Ross} \& {Fabian}}{{Ross} \&
  {Fabian}}{1993}]{1993MNRAS.261...74R}
{Ross} R.~R.,  {Fabian} A.~C.,  1993, \mn@doi [\mnras]
  {10.1093/mnras/261.1.74}, \href
  {http://adsabs.harvard.edu/abs/1993MNRAS.261...74R} {261, 74}

\bibitem[\protect\citeauthoryear{{Rybicki} \& {Lightman}}{{Rybicki} \&
  {Lightman}}{1979}]{1979rpa..book.....R}
{Rybicki} G.~B.,  {Lightman} A.~P.,  1979, {Radiative processes in
  astrophysics}

\bibitem[\protect\citeauthoryear{{Shakura} \& {Sunyaev}}{{Shakura} \&
  {Sunyaev}}{1973}]{1973A&A....24..337S}
{Shakura} N.~I.,  {Sunyaev} R.~A.,  1973, \aap, \href
  {http://adsabs.harvard.edu/abs/1973A%26A....24..337S} {24, 337}

\bibitem[\protect\citeauthoryear{{Stern}, {Poutanen}, {Svensson}, {Sikora}  \&
  {Begelman}}{{Stern} et~al.}{1995}]{1995ApJ...449L..13S}
{Stern} B.~E.,  {Poutanen} J.,  {Svensson} R.,  {Sikora} M.,   {Begelman}
  M.~C.,  1995, \mn@doi [\apjl] {10.1086/309617}, \href
  {http://adsabs.harvard.edu/abs/1995ApJ...449L..13S} {449, L13}

\bibitem[\protect\citeauthoryear{{Tazaki}, {Ueda}, {Ishino}, {Eguchi}, {Isobe},
  {Terashima}  \& {Mushotzky}}{{Tazaki} et~al.}{2010}]{2010ApJ...721.1340T}
{Tazaki} F.,  {Ueda} Y.,  {Ishino} Y.,  {Eguchi} S.,  {Isobe} N.,  {Terashima}
  Y.,   {Mushotzky} R.~F.,  2010, \mn@doi [\apj]
  {10.1088/0004-637X/721/2/1340}, \href
  {http://adsabs.harvard.edu/abs/2010ApJ...721.1340T} {721, 1340}

\bibitem[\protect\citeauthoryear{{Tazaki}, {Ueda}, {Terashima}  \&
  {Mushotzky}}{{Tazaki} et~al.}{2011}]{2011ApJ...738...70T}
{Tazaki} F.,  {Ueda} Y.,  {Terashima} Y.,   {Mushotzky} R.~F.,  2011, \mn@doi
  [\apj] {10.1088/0004-637X/738/1/70}, \href
  {http://adsabs.harvard.edu/abs/2011ApJ...738...70T} {738, 70}

\bibitem[\protect\citeauthoryear{{Tortosa} et~al.,}{{Tortosa}
  et~al.}{2016}]{2016arXiv161205871T}
{Tortosa} A.,  et~al., 2016, preprint, \href
  {http://adsabs.harvard.edu/abs/2016arXiv161205871T} {} (\mn@eprint {arXiv}
  {1612.05871})

\bibitem[\protect\citeauthoryear{Tortosa et~al.,}{Tortosa
  et~al.}{2018a}]{2018MNRAS.473.3104T}
Tortosa A.,  et~al., 2018a, \mn@doi [Monthly Notices of the Royal Astronomical
  Society] {10.1093/mnras/stx2457}, 473, 3104

\bibitem[\protect\citeauthoryear{Tortosa, Bianchi, Marinucci, Matt  \&
  Petrucci}{Tortosa et~al.}{2018b}]{2018A&A...614A..37T}
Tortosa A.,  Bianchi S.,  Marinucci A.,  Matt G.,   Petrucci P.~O.,  2018b,
  \mn@doi [Astron. Astrophys.] {10.1051/0004-6361/201732382}, 614, A37

\bibitem[\protect\citeauthoryear{{Urry} \& {Padovani}}{{Urry} \&
  {Padovani}}{1995}]{1995PASP..107..803U}
{Urry} C.~M.,  {Padovani} P.,  1995, \mn@doi [\pasp] {10.1086/133630}, \href
  {http://adsabs.harvard.edu/abs/1995PASP..107..803U} {107, 803}

\bibitem[\protect\citeauthoryear{{Ursini} et~al.,}{{Ursini}
  et~al.}{2015}]{2015MNRAS.452.3266U}
{Ursini} F.,  et~al., 2015, \mn@doi [\mnras] {10.1093/mnras/stv1527}, \href
  {http://adsabs.harvard.edu/abs/2015MNRAS.452.3266U} {452, 3266}

\bibitem[\protect\citeauthoryear{{Ursini} et~al.,}{{Ursini}
  et~al.}{2016}]{2016MNRAS.463..382U}
{Ursini} F.,  et~al., 2016, \mn@doi [\mnras] {10.1093/mnras/stw2022}, \href
  {http://adsabs.harvard.edu/abs/2016MNRAS.463..382U} {463, 382}

\bibitem[\protect\citeauthoryear{{Vasudevan} \& {Fabian}}{{Vasudevan} \&
  {Fabian}}{2007}]{2007MNRAS.381.1235V}
{Vasudevan} R.~V.,  {Fabian} A.~C.,  2007, \mn@doi [\mnras]
  {10.1111/j.1365-2966.2007.12328.x}, \href
  {http://adsabs.harvard.edu/abs/2007MNRAS.381.1235V} {381, 1235}

\bibitem[\protect\citeauthoryear{{Vasudevan} \& {Fabian}}{{Vasudevan} \&
  {Fabian}}{2009}]{2009MNRAS.392.1124V}
{Vasudevan} R.~V.,  {Fabian} A.~C.,  2009, \mn@doi [\mnras]
  {10.1111/j.1365-2966.2008.14108.x}, \href
  {http://adsabs.harvard.edu/abs/2009MNRAS.392.1124V} {392, 1124}

\bibitem[\protect\citeauthoryear{{Vasudevan}, {Brandt}, {Mushotzky}, {Winter},
  {Baumgartner}, {Shimizu}, {Schneider}  \& {Nousek}}{{Vasudevan}
  et~al.}{2013a}]{2013ApJ...763..111V}
{Vasudevan} R.~V.,  {Brandt} W.~N.,  {Mushotzky} R.~F.,  {Winter} L.~M.,
  {Baumgartner} W.~H.,  {Shimizu} T.~T.,  {Schneider} D.~P.,   {Nousek} J.,
  2013a, \mn@doi [\apj] {10.1088/0004-637X/763/2/111}, \href
  {http://adsabs.harvard.edu/abs/2013ApJ...763..111V} {763, 111}

\bibitem[\protect\citeauthoryear{{Vasudevan}, {Mushotzky}  \&
  {Gandhi}}{{Vasudevan} et~al.}{2013b}]{2013ApJ...770L..37V}
{Vasudevan} R.~V.,  {Mushotzky} R.~F.,   {Gandhi} P.,  2013b, \mn@doi [\apjl]
  {10.1088/2041-8205/770/2/L37}, \href
  {http://adsabs.harvard.edu/abs/2013ApJ...770L..37V} {770, L37}

\bibitem[\protect\citeauthoryear{{Vaughan}, {Edelson}, {Warwick}  \&
  {Uttley}}{{Vaughan} et~al.}{2003}]{2003MNRAS.345.1271V}
{Vaughan} S.,  {Edelson} R.,  {Warwick} R.~S.,   {Uttley} P.,  2003, \mn@doi
  [\mnras] {10.1046/j.1365-2966.2003.07042.x}, \href
  {http://adsabs.harvard.edu/abs/2003MNRAS.345.1271V} {345, 1271}

\bibitem[\protect\citeauthoryear{{V{\'e}ron-Cetty} \&
  {V{\'e}ron}}{{V{\'e}ron-Cetty} \& {V{\'e}ron}}{2010}]{2010A&A...518A..10V}
{V{\'e}ron-Cetty} M.-P.,  {V{\'e}ron} P.,  2010, \mn@doi [\aap]
  {10.1051/0004-6361/201014188}, \href
  {http://adsabs.harvard.edu/abs/2010A%26A...518A..10V} {518, A10}

\bibitem[\protect\citeauthoryear{{Wilms}, {Allen}  \& {McCray}}{{Wilms}
  et~al.}{2000}]{2000ApJ...542..914W}
{Wilms} J.,  {Allen} A.,   {McCray} R.,  2000, \mn@doi [\apj] {10.1086/317016},
  \href {http://adsabs.harvard.edu/abs/2000ApJ...542..914W} {542, 914}

\bibitem[\protect\citeauthoryear{{Woo} \& {Urry}}{{Woo} \&
  {Urry}}{2002}]{2002ApJ...579..530W}
{Woo} J.-H.,  {Urry} C.~M.,  2002, \mn@doi [\apj] {10.1086/342878}, \href
  {http://adsabs.harvard.edu/abs/2002ApJ...579..530W} {579, 530}

\bibitem[\protect\citeauthoryear{{Zdziarski}, {Johnson}  \&
  {Magdziarz}}{{Zdziarski} et~al.}{1996}]{1996MNRAS.283..193Z}
{Zdziarski} A.~A.,  {Johnson} W.~N.,   {Magdziarz} P.,  1996, \mn@doi [\mnras]
  {10.1093/mnras/283.1.193}, \href
  {http://adsabs.harvard.edu/abs/1996MNRAS.283..193Z} {283, 193}

\bibitem[\protect\citeauthoryear{{Zdziarski}, {Poutanen}  \&
  {Johnson}}{{Zdziarski} et~al.}{2000}]{2000ApJ...542..703Z}
{Zdziarski} A.~A.,  {Poutanen} J.,   {Johnson} W.~N.,  2000, \mn@doi [\apj]
  {10.1086/317046}, \href {http://adsabs.harvard.edu/abs/2000ApJ...542..703Z}
  {542, 703}

\bibitem[\protect\citeauthoryear{{Zhang}, {Wang}  \& {Zhu}}{{Zhang}
  et~al.}{2018}]{2018ApJ...863...71Z}
{Zhang} J.-X.,  {Wang} J.-X.,   {Zhu} F.-F.,  2018, \mn@doi [\apj]
  {10.3847/1538-4357/aacf92}, \href
  {http://adsabs.harvard.edu/abs/2018ApJ...863...71Z} {863, 71}

\bibitem[\protect\citeauthoryear{{Zoghbi} et~al.,}{{Zoghbi}
  et~al.}{2017}]{2017ApJ...836....2Z}
{Zoghbi} A.,  et~al., 2017, \mn@doi [\apj] {10.3847/1538-4357/aa582c}, \href
  {http://adsabs.harvard.edu/abs/2017ApJ...836....2Z} {836, 2}

\bibitem[\protect\citeauthoryear{{Zycki} \& {Czerny}}{{Zycki} \&
  {Czerny}}{1994}]{1994MNRAS.266..653Z}
{Zycki} P.~T.,  {Czerny} B.,  1994, \mn@doi [\mnras] {10.1093/mnras/266.3.653},
  \href {http://adsabs.harvard.edu/abs/1994MNRAS.266..653Z} {266, 653}

\makeatother
\end{thebibliography}

\label{lastpage}
\end{document}